\documentclass[pre,twocolumn,superscriptaddress,floatfix]{revtex4}
\usepackage{epsfig}
\usepackage{subfigure}
\usepackage{graphicx}
\usepackage{amssymb}
\usepackage{epstopdf}
\usepackage{color}

\begin{document}

\title{Understanding tetrahedral liquids through patchy colloids}

\author{Ivan Saika-Voivod}
\affiliation{Department of Physics and Physical Oceanography,
Memorial University of Newfoundland, St. John's, NL, A1B 3X7, Canada}

\author{Frank Smallenburg}
\affiliation{Dipartimento di Fisica, Sapienza Universit\`a di Roma, Piazzale Aldo Moro 5, 00185 Roma, Italy}

\author{Francesco Sciortino}
\affiliation{Dipartimento di Fisica, Sapienza Universit\`a di Roma, Piazzale Aldo Moro 5, 00185 Roma, Italy}

\date{\today}

\begin{abstract}

We investigate the structural properties of a simple model for tetrahedral patchy colloids in which
the patch width and the patch range can be tuned independently.  
For wide bond angles, a fully bonded network can be generated by standard Monte Carlo or molecular dynamics simulations of the model, providing a neat method for generating  defect-free random tetrahedral networks. This offers the possibility
of focusing on the role of the patch angular width on the structure of the fully bonded network.  The analysis of the fully bonded configurations as a function of the bonding angle shows how the bonding angle controls the system compressibility, the strength of the pre-peak in the structure factor and ring size distribution. 
 Comparison with  models of liquid  water and silica allows us to find the best mapping  between these continuous potentials and the colloidal one. 
 Building on previous studies focused on the connection between angular range and
crystallization, the mapping makes it possible to shed
 new light on the  glass-forming ability of network-forming tetrahedral liquids.


\end{abstract}

\maketitle

\section{Introduction}


Cooling a liquid below its melting temperature will result in either crystallization or vitrification~\cite{kob-binder-book}. Understanding on a basic
level why one alternative is predominantly chosen over the other in different systems remains fundamentally an open problem.
This is particularly acute in tetrahedral network-forming liquids such as water and silica.  In both liquids, corner-sharing 
tetrahedral units form a random network that fills space.  However, silica is the prototypical glass former, while water crystallizes
readily.

Recent advances in the synthesis of colloidal particles have brought about the patchy particle~\cite{Pawar:2010ig,BianchiPCCP}.  
Of particular relevance as models of networked liquids are colloids 
with sticky spots that provide strongly directional interactions with fixed valence~\cite{Wang:2012gd,biffi,Granick}.  To better understand the physics of self-assembly in these systems, theorists have been studying  colloidal particles  with simple models that incorporate information concerning the valence. These models can be
considered as an evolution of primitive models for associated liquids~\cite{Vlc04a,Nez05a,Kolafa:1987wk}.  
Particularly relevant for our investigation is the Kern-Frenkel (KF) model~\cite{Kern:2003ie}, a model 
 which has its origin in the work of Bol~\cite{bol} and that offers a very clean separation between 
the angular width and radial extent of the attractive interaction between patches.

Recent systematic simulation studies of the KF model with tetrahedrally arranged patches have uncovered the dependence
of both the driving force for nucleation, namely the chemical potential difference between crystal and liquid, and the free energy barriers
to nucleation on the width of the patches~\cite{Romano:2010bl,Romano:2011jm,SaikaVoivod:2011co}.  More recently, the KF model was extended to retain its tetravalent character even in the 
case where patches are wide enough to accommodate more than one bond. 
Extending the model to guarantee the single-bond-per-patch condition regardless of
patch width, under appropriate conditions, results in the liquid
retaining thermodynamic stability down to zero temperature, avoiding the crystal phase altogether~\cite{Smallenburg:2013cq}.

The relative simplicity of the KF model allows us to understand its thermodynamic properties through bonding entropy in
geometric terms, at least for the crystal.  For the liquid, the computational efficiency with which the model can be simulated
allows for detailed calculations of contributions to the free energy down to states at or near the ground state, which has a
well defined energy.
Beyond being a model for patchy colloids, the KF model provides a potential coarse grained description of
bond flexibility in more general network-forming systems.  One requires a way of mapping systems onto the KF model with 
particular patch widths.

In particular, if one is able to provide such a map for different network-forming liquids, then one can unlock the insights
we have gained from the patch width dependence of crystallization in KF to inform our understanding of why, for example,
water freezes while silica does not.

In the present work, we compare the networks formed in BKS silica~\cite{vanBeest:1990tt}, 
ST2 water~\cite{Stillinger:1974fh}, TIP4P/2005 water~\cite{Abascal:2005ka}, 
mW water~\cite{Molinero:2009kc} and Stillinger-Weber (SW) silicon~\cite{Stillinger:1985vx}.
We do so always at the optimal network-forming density for each system~\cite{simone}, thus eliminating density as a parameter, 
and find markers of network formation that allow us to make a mapping of these systems onto KF models 
with different patch widths.  
We focus on the best available network configurations across the systems.  For the KF model, we utilize only fully bonded networks.

The models represent different classes in the way that tetrahedral geometry is enforced.  For KF, mW and SW, the geometry is enforced at the level of bonding between particles, whether through the placement of patches, or through a three-body 
interaction term.  ST2 and TIP4P/2005 have tetrahedrality built into the rigid shape of the water molecule through charge sites, 
which in turn direct hydrogen bonding.  The BKS model is a binary mixture, with bridging between Si ions mediated by relatively large and soft oxygen
ions that arrange themselves tetrahedrally, essentially through stoichiometry and steric repulsion.  All of these systems can be understood in terms of the random network 
model~\cite{Zachariasen:1932tn,Evans:1966vk,Bell:1966ww,Sceats:1979em,Rice:1981vz,Henn:1989ua,Polk:73wr}, 
although the perturbations required to 
quantitatively account for the particularities of each system are potentially qualitatively different.
Given these differences, it is not obvious that a mapping to the KF model via the patch width, a single parameter,
is a realistic endeavor from the outset.

Mapping of various network-forming systems to other models where bond flexibility can be tuned, such as 
the mW--SW family of models, could prove a worthwhile pursuit. The work of Molinero and coworkers~\cite{Molinero:2006hk}
on the dependence of the location of the liquid-liquid critical point in the model on the strength of the three-body
interaction term lays the groundwork for such a study.  However, the extent to which crystallization is understood in the KF model
makes it an ideal candidate for unlocking insights into other network-forming liquids.

Additionally, we explore the ability of fully-bonded KF networks to generate amorphous ground states for silica and water.
A long-standing issue within the scientific community is the development of algorithms that can generate ideal random
tetrahedral networks~\cite{Wooten:1985vx,Barkema:2000ut,Mousseau:2004fk}.  
Thus, it is of interest to see how perfection in the KF model transfers, from an energetic perspective, 
to models for such emblematic network formers such as silica and water.

\section{Methods}

In this work, we compare the properties of the best available tetrahedral networks from several model liquids:
the KF model for tetravalent patchy colloids,
the BKS model of silica, the ST2 model for water, the mW water model and the SW
model of silicon.  To facilitate comparison, structural quantities for all models are reported at or near the optimal density~\cite{simone} or pressure for
the formation of a bonded network.
Generally, we identify this optimal condition with a local minimum in low temperature isotherms of potential
energy.  Further, all structural quantities are reported for configurations that are first quenched through a conjugate-gradient (CG) algorithm to a local 
minimum in the potential energy, i.e., to a so-called inherent structure (IS)~\cite{Sti88a,Sti95a,Stathmech}.  This eliminates disorder due to vibrations.  

\subsection{KF model}

The KF model we consider consists of hard sphere particles of diameter $\sigma$, each decorated with four tetrahedrally arranged patches.
Each patch is defined by a cone with apex at the particle centre and aperture $2\theta$, where $\theta$ is the angle between the cone axis
and a generatrix.
Particles form bonds of energy $-\epsilon$ when patches overlap.  Overlap occurs when the center-to-center  
distance between two particles is less than $(1+\delta)\sigma$, with $\delta=0.12$, and the line segment connecting particle centres passes through 
both patches, i.e., when the angle between the line segment and each of the patch axes is less than $\theta$.  See Ref.~\cite{Romano:2010bl} for details.

For sufficiently wide patches, i.e., for $2\sin{\theta}>(1+\delta)^{-1}$ (in our case $\theta > 26.5^\circ$, or $\cos{\theta}<0.895$), it becomes possible
for a patch on one particle to form bonds simultaneously with more than one particle.  To avoid this scenario, we enforce single bonding per patch
through the extension of the KF model presented in Ref.~\cite{Smallenburg:2013cq} (and associated supplementary information), where two of us
showed that for sufficiently wide patches over an appropriate range in density, the liquid state is more stable than the crystal down to $T=0$.
Briefly, the single-bond-per-patch condition is enforced by allowing bonds to form between overlapping patches subject to a probability based
on the Boltzmann distribution.  When more than one overlap occurs with a patch, only one of the overlaps, selected at random, is allowed to form
a bond.  Thus, for all patch widths the ground state energy per particle is $-\epsilon/2$.  For $\cos{\theta}\ge0.895$, where the narrowness of the patches
naturally restricts patches to only share a single bond, the extended KF model maps onto the original KF model with a rescaled $\epsilon$, or equivalently,
at a rescaled $T$.  At low $T$, the two models become equivalent.

The single-bond-per-patch condition is implemented in the event-driven molecular dynamics (EDMD) simulations of the model by introducing an
event that randomly selects a bond to be re-evaluated according to the Boltzmann criterion employed to define bonds.  The events are scheduled
at random according to an exponential distribution with an average rate that is sufficiently high to ensure that the system dynamics are invariant with 
respect to the re-evaluation procedure.  See Ref.~\cite{Smallenburg:2013cq} again for details. 

We also carry out Monte Carlo (MC) simulations of the original KF model (without enforced single bonding), 
as in Ref.~\cite{Romano:2010bl}, for  $\cos{\theta}\ge0.895$, as well as for the modified KF model with single bonding enforced. 
We find that MC is more efficient at equilibrating the systems at low $T$ as the patches become narrow.
Additionally, these simulations serve as a consistency check on the EDMD simulations. 

Regardless of the simulation method, we report structural quantities only for fully-bonded KF configurations, i.e., perfect networks
at the ground state energy, at a density of $\rho \sigma^3 = 0.57$.  Obtaining perfect networks becomes progressively more
computationally challenging as $\cos{\theta}$ increases, and we are not able to obtain perfect liquid configurations beyond
$\cos{\theta}=0.92$.  For $\cos{\theta}=0.85$, 0.87, 0.895 and 0.92, we obtain only a single perfect configuration;
these perfect narrow-patch configurations are obtained through MC.

\subsection{BKS silica}

For BKS silica, we simulate 444 SiO$_2$
units in a cubic box of length $2.65761$~nm, i.e.,  at a 
constant density of $2.36$~g/cm$^3$.  At this density at low $T$, the tetrahedral
network is well formed~\cite{Horbach:1999ib}, and this density is in fact near the minimum in potential energy of 
BKS silica as a function of density at $T=3000$~K~\cite{SaikaVoivod:2004jf,SaikaVoivod:2005vt}.  
Constant $T$ molecular dynamics simulations from $T$=3000 to 2400~K are carried out with a time step of 1~fs 
using Gromacs software version 
4.5.5~\cite{Berendsen:1995tn,Lindahl:2001bm,vanderSpoel:2005hz,Hess:2008db} 
employing the 
Nos\'e-Hoover thermostat with a time constant of 1ps and a radial cutoff of 1~nm for all real space pair interactions.  
Coulomb interactions are handled with the particle mesh Ewald (PME)
algorithm with a fourier spacing of 0.1nm and interpolation of order four (cubic). 
 We add to the BKS potential a short range interaction, described
in~\cite{SaikaVoivod:2004jf}, to prevent the system from exploring the unphysical attraction occurring at small distances.
At 2400~K, we simulate for  
2.4~$\mu$s as the system dynamics are quite slow, with Si ions diffusing a root mean square distance of approximately
0.35~nm (just over one average Si-Si distance) in 1~$\mu$s.  We use configurations from $T=2400$~K for structural analysis.

In order to facilitate comparison of the IS energy $e_{\rm IS}$
with previous work, in performing the quench we employ the slightly modified form of BKS employed in~\cite{SaikaVoivod:2004jf}, 
in which the real space part of the potential tapers smoothly starting at 0.77476~nm from the original BKS value to zero at 1~nm.  
This procedure recovers the previously 
reported IS energies in Ref.~\cite{SaikaVoivod:2004jf}  for the $T$ overlapping with this work ($2700$~K to $3000$~K),
as shown in Fig.~\ref{figenergy}(d). 

\subsection{Water and silicon models}

For our simulations of the TIP4P/2005 model of water, we also use Gromacs v4.5.5 to carry out a series of 
$NVT$ simulations, varying the number of particles from 231 to 309 and keeping the cubic box fixed at 
$V=(2$~nm$)^3$, thus varying density from approximately 0.86 to 1.16 g/cm$^3$.  
Following Ref.~\cite{Abascal:2010dw}, we employ a time step of 2~fs, a Nos\'e-Hoover thermostat with a time constant of 1~ps and a real space potential cutoff of
0.85~nm. For Coulomb terms, we use the PME algorithm with  fourier spacing of 0.1~nm and interpolation order four (cubic).
Initial configurations are equilibrated at 235~K before running at $T=193$~K.  At the dynamically slowest state point (lowest density), the potential energy
reaches a steady state after 10~$\mu$s.  For the ensuing 10~$\mu$s, molecules diffuse a root mean square distance
of 0.43~nm.  While this is not sufficient to obtain very precise averages at low density, it is sufficient to discern a minimum in the energy as a function of density near 0.92~g/cm$^3$. For structural analysis, we therefore harvest configurations from our simulation containing 245 molecules, which 
corresponds to this optimal density.

Configurations for ST2 water (employing the reaction field treatment of electrostatic interactions) 
are taken from a density of 0.83~g/cm$^3$ (222 water molecules in a box 
of length 2~nm) at $T=235$~K. 
To obtain these low $T$ configurations, we extend the successive umbrella sampling grand canonical Monte Carlo simulations of ST2 described in Ref.~\cite{Sciortino:2011gf}.

To extend our analysis to another class of tetrahedral liquid models, we perform a less thorough investigation of two members of the SW silica 
model family, that have different values of the parameter controlling the strength of the three-body, tetrahedrality-enforcing interaction term.
We thank colleagues Vishwas Vashist and Srikanth Sastry for providing us with quenched configurations of the SW model with $10000$ particles at $T=1196$~K and $P=-1.88$GPa.  The state point is above the liquid-liquid critical temperature for that model, but below the line of compressibility maxima and in the tetrahedral network 
regime~\cite{Vasisht:2011ch}.  We are also grateful to colleagues Dr.~Flavio Romano and Dr.~John Russo for providing us, 
in the course of their current study of the model, 
with quenched configurations of 686 mW water molecules at $P=0$ and $T=171.3$~K, a state point below the transition temperature to 
the low density liquid (LDL) as presented in~\cite{Molinero:2009kc}, but at which crystallization also occurs.  

\subsection{Structural quantities}

We report the structure factor $S(q)$, defined as 
$
\left< \rho_{\vec{q}} \rho_{-\vec{q}} \right>/N,
$
where
$
\rho_{\vec{q}}=\sum_{i=1}^N \exp{(-\vec{q} \cdot \vec{r}_i) },
$
$\left< \dots \right>$ denotes an ensemble average over reciprocal space vectors $\vec{q}$ having magnitude $q$ and
$N$ is the number of {\it node} particles in the systems.  A node particle is simply the particle that is at the centre of a tetrahedron in 
the corner sharing tetrahedral network: for the KF, mW and SW models, each particle is a node
particle; for silica, the Si ions are node particles, while O atoms are the node particles for ST2 and TIP4P/2005 water.  We also calculate the radial distribution 
function $g(r)$ for node particles.

The other key structural quantity we present is the distribution of the angle $\phi$, defined by the vectors emanating from a node particle to
two of its neighboring node particles (the Si-Si-Si angle for BKS and SW, O-O-O angle for water). For the continuous potential models, we define neighbors of a particle as those that lie within a radius of $r_{\rm cut}$,
determined from the minimum between first and second neighbor peaks in $g(r)$.  Values of $r_{\rm cut}$ are given in 
Table~\ref{tab1}. For the KF model, neighbors are defined as particles
sharing a bond.

Additionally, we report on the distribution of the sizes of minimal closed rings of neighboring node particles, with neighbors defined as above for
the bond angle distribution. The algorithm for determining ring statistics is from Ref.~\cite{Yuan:2002tt}.

\section{Results}

\subsection{$S(q)$ and $g(r)$}

\begin{figure}
\hbox to \hsize{\epsfxsize=1\hsize\epsfbox{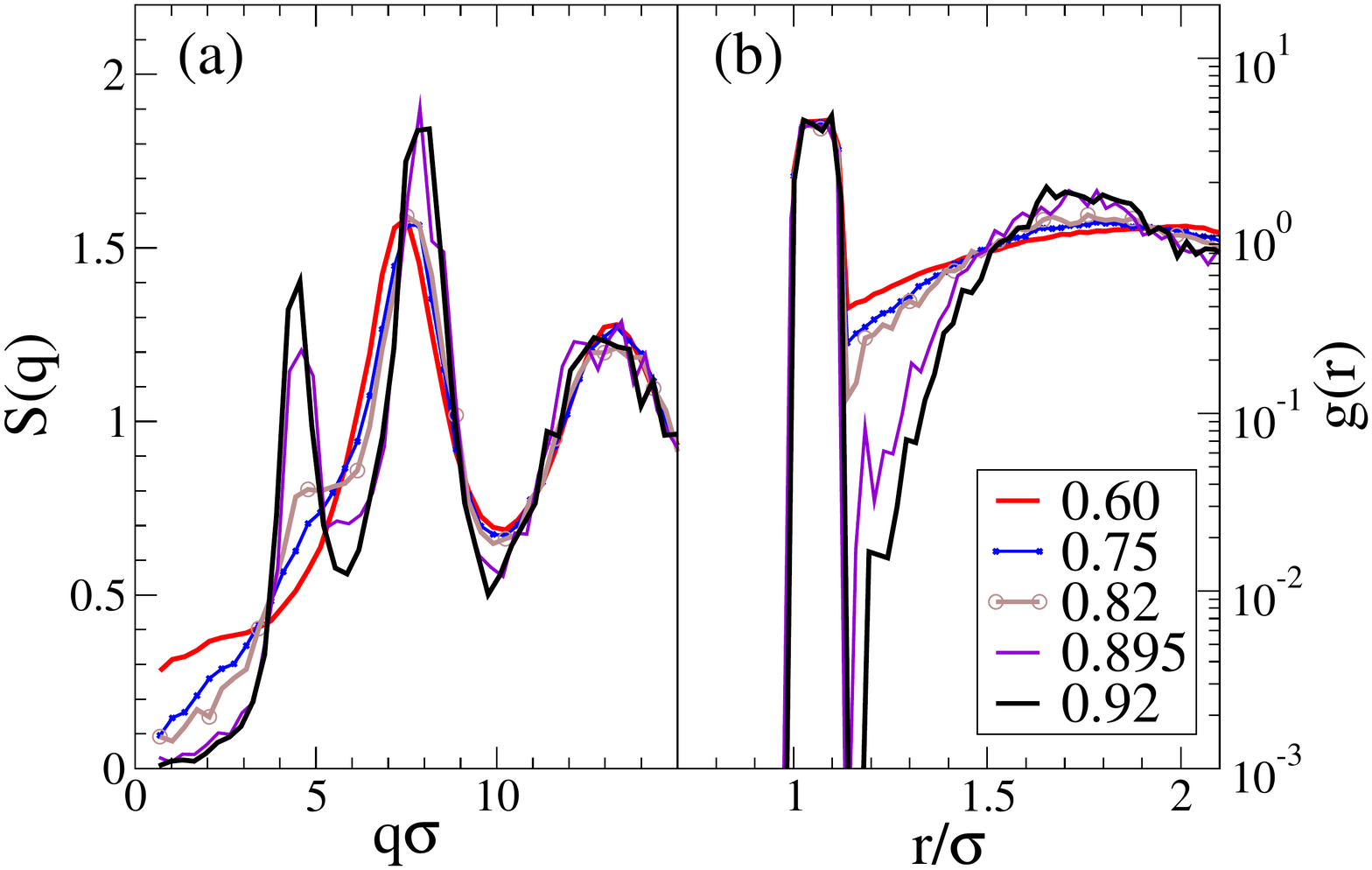}}
\hbox to \hsize{\epsfxsize=1\hsize\epsfbox{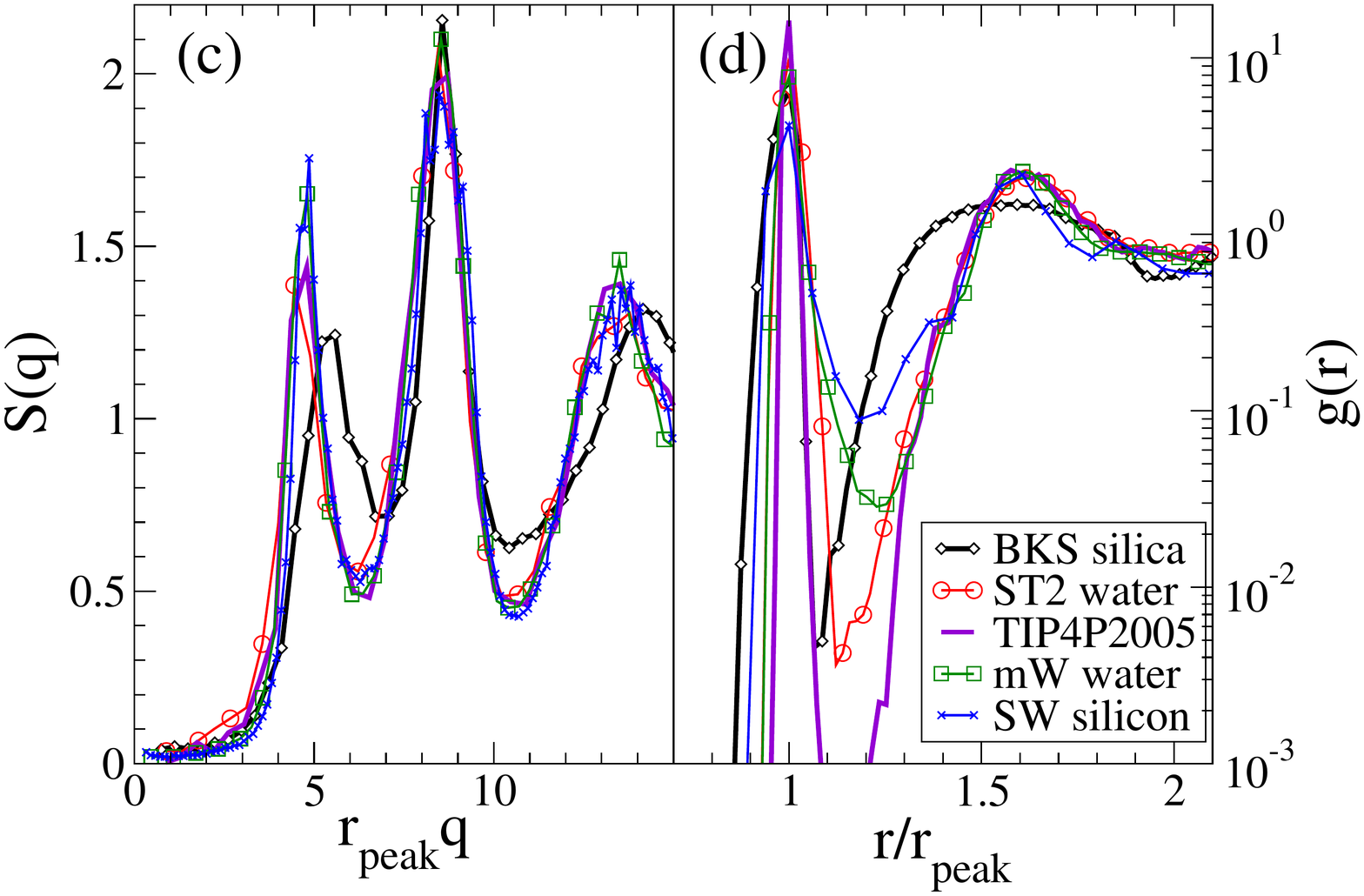}}
\caption{Structure factor and radial distribution functions.  
Upper panels give (a) $S(q)$ and (b)  $g(r)$ for the KF model at $\rho\sigma^3=0.57$.  Legend in (b) gives values of $\cos{\theta}$.
There is a progressively higher network peak in $S(q)$ (near q$\sigma\approx5$) as patches become narrower and a
supporting trend in $g(r)$, with progressively better defined neighbor coordination shells.  
Lower panels give (c) $S(q)$ and (d) $g(r)$ for the continuous models, 
for which we rescale distance with $r_{\rm peak}$, the position of the first peak in $g(r)$.  
}
\label{figSq}
\end{figure}

\begin{table}
\begin{tabular}{|c|c|c|c|c|c|}
\hline
Model & $r_{\rm peak}$  & $\rho_{\rm eff}^*$ & $r_{\rm cut}$ & $\bar{n}_b$  &  $f_{\rm def}$ \\
\hline
KF & 1.06$\sigma$ & 0.68 & n.a. & 4 & 0 \\
BKS & 0.315~nm & 0.74 & 0.329~nm & 4.002 & 0.0024 \\
ST2 & 0.283~nm & 0.63 & 0.320~nm & 3.988 & 0.018 \\
TIP4P/2005 & 0.2775~nm & 0.65 & 0.315~nm & 3.996 & 0.0046 \\
mW & 1.143$\sigma$ & 0.66 & 1.40$\sigma$ & 3.63 & 0.327 \\
SW & 1.165$\sigma$ & 0.70 & 1.41$\sigma$ & 3.50 & 0.475 \\
\hline
\end{tabular}
\caption{Summary of parameters characterizing the systems studied: $r_{\rm peak}$ is the position of the first peak in $g(r)$;
$\rho_{\rm eff}^*$ is the value of the reduced density $\rho \, r_{\rm peak}^3$ 
at which structural analysis is reported;
$r_{\rm cut}$ is the position of the first minimum in $g(r)$, and is used to determine neighbors;
 $\bar{n}_b$ is the average number of neighbors;
 $f_{\rm def}$ is the fraction of particles that do not have four neighbors. 
For mW $\sigma$=0.23925~nm and  for SW $\sigma$=0.20951~nm.}
\label{tab1}
\end{table}

In Figs.~\ref{figSq}(a) and (b), we present $S(q)$
and $g(r)$ for the tetrahedral KF model with patch angles ranging
from $\cos\theta=0.60$ ($\theta=53.1^\circ$) to $\cos\theta$=0.92 ($\theta=23.1^\circ$).
The perfectly bonded configurations from which the curves are derived are obtained from low $T$ simulations which
sample the energetic ground state.  Beyond $\cos\theta$=0.82, the computational effort required to sample such states
grows considerably. 
As $\cos\theta$ increases, i.e., as the patch angle decreases, the peak characteristic of a structured network develops~\cite{Elliott:2001tp}.
First a shoulder appears, and by $\cos\theta=0.82$ a peak in the form of a local maximum is established.
Beyond $\cos\theta=0.82$, the position $q_1$ of this network peak does not change significantly, while its height $S(q_1)$ grows 
(as does the height of the main peak at $\sigma q_2\approx8$). 
Accompanying the emergence and growth of the network peak in $S(q)$ is 
 a significant decrease of the system compressibility, as evidenced by the
approach toward zero of $S(0)$. Small values of $S(0)$ are commonly 
understood to signify
high degrees of hyperuniformity~\cite{torquatopnas}. It is interesting to observe in passing that
for the case of $\cos(\theta)=0.92$,  despite our inability to accurately evaluate $S(0)$ 
 with the present system size,  $S(0) \approx 0.005$, a value  smaller than
the  computationally determined lower bound recently suggested by de Graff and Thorpe~\cite{thorpe}
based on studies of continuous random network models. The observed value is comparable to the experimentally determined value for annealed a-Si~\cite{torquatopnas}.  Data also show a progressive sharpening of the  second neighbor peak in $g(r)$ associated with the growth of the network peak in $S(q)$.  Not surprisingly, the region between first and second neighbor shells becomes more and more depleted of particles.
The first peak in $g(r)$ drops off very sharply to a minimum 
just outside the bonding cutoff distance of $1+\delta=1.12$.

In Figs.~\ref{figSq}(c) and (d), we compare $S(q)$ and $g(r)$ for the various continuous tetrahedral liquid models to each other.
To facilitate comparison, we rescale $q$ and $r$ with the position $r_{\rm peak}$ of the first peak in $g(r)$.  The values of
$r_{\rm peak}$ used in the rescaling are given in Table~\ref{tab1}.
An inspection of $S(q_1)$ suggests a reasonable similarity between ST2 and TIP4P/2005 with $\cos\theta=0.92$. 
SW and mW, while similar to each other, have a significantly higher value of $S(q_1)$ compared to other models. 
BKS, however, not only
has a significantly lower peak at $q_1$, but also has a value of $q_1$ larger than the common value of $q_1$ 
shared by all the other models. Thus, despite having the largest value of $S(q_2)$, the network in BKS is distinct from the
other models. In terms of $g(r)$, the first neighbor peaks of $\cos\theta=0.92$, KF, ST2 and TIP4P/2005 are similar in their sharp fall-off
at a similar reduced distance.  However, there is a significant difference in the position and width of the second peak,  
particularly for that of BKS, which is both closer and wider.  Thus, it is not surprising that for BKS 
$q_1$ is larger and $S(q_1)$ smaller than for the other models.

We also note that the depth of the minimum at $r_{\rm cut}$ in $g(r)$, while generally a good indicator of ordering in the liquid, 
does not uniquely define the number of defects present in the tetrahedral liquid.  
For example, 
once we define the neighbors of a particle as those lying
with a distance $r_{\rm cut}$ of that particle, then a direct calculation of the average number of neighbors 
reveals that while $g(r_{\rm cut})$ for TIP4P/2005 is more than an order of magnitude lower than for BKS, 
the deviation of the number of neighbors from four is about the same for both models.
The average number of neighbors and the fraction of defects for the various models are given in Table~\ref{tab1}.  
The fraction
of defects is simply the fraction of node particles that do not have four neighbors.
Generally, mW and SW show a relatively large number of defects, and are also both known to crystallize spontaneously
in simulation without great difficulty~\cite{Vasisht:2011ch,Molinero:2009kc}.

\subsection{Energy as a function of density}

\begin{figure}
\hbox to \hsize{\epsfxsize=1\hsize\epsfbox{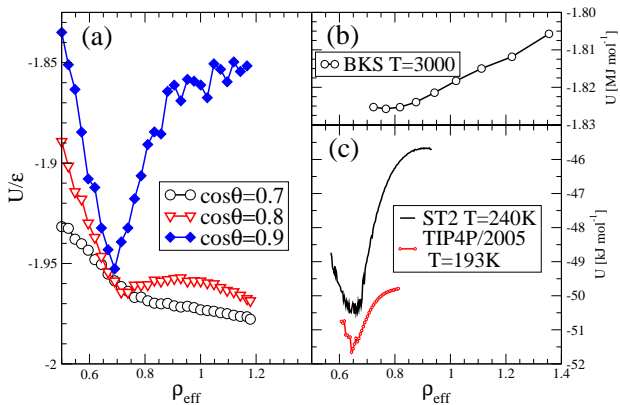}}
\caption{Potential energy isotherm for (a) the single-bond-per-patch KF model at $k_B T/\epsilon = 1/9$ (b) BKS silica at $T=3000$~K taken from Ref.~\cite{SaikaVoivod:2004jf}
and (c) ST2 at $T=240~K$ taken from Ref.~\cite{Sciortino:2011gf} and TIP4P/2005 at $T=193$, for which the data are shifted up by 4.5 kJ/mol to
facilitate comparison with the ST2 model. 
}
\label{figurho}
\end{figure}

The scaling of distance with $r_{\rm peak}$ allows us to compare optimal network densities by defining a reduced 
number density $\rho_{\rm eff}=\rho \, r_{\rm peak}^3$.  Table~\ref{tab1} shows the value of the reduced density 
$\rho^*_{\rm eff}$ used for structural analysis of the network for the models we study.

Fig.~\ref{figurho}(a) shows
the potential energy $U$ of the KF model with the single-bond-per-patch condition enforced at $k_B T/\epsilon = 1/9$ for
$\cos\theta =0.70$, 0.80 and 0.90 as a function of $\rho_{\rm eff}$.
At $\cos\theta =0.90$, we see a well defined minimum at $\rho_{\rm eff}=0.69$ ($\rho \sigma^3=0.58$).
At $\cos\theta =0.80$, there is a shallow minimum at a slightly higher value of $\rho_{\rm eff}$, 
while for $\cos\theta =0.70$ there is only a kink in the curve near $\rho_{\rm eff}=0.8$.  The appearance of a
minimum in $U(\rho)$ is thus concurrent with the appearance of the network peak in $S(q)$.  And as the patch width
becomes narrower, this minimum shifts to lower density.  
We note that for $\cos\theta =0.92$ and $\delta=0.12$, the density range for single phase stability for the diamond structure is narrow
and occurs near $\rho \sigma^3=0.57$ or $\rho_{\rm eff}=0.68$.
For reference, the number density of the diamond cubic structure for touching hard spheres of unit diameter is $3\sqrt{3}/8\approx0.65$


$U(\rho)$ at low $T$ for the molecular water models in Fig.~\ref{figurho}(c) show minima at  $\rho_{\rm eff}\approx0.65$, and while the
rise in energy with increasing density is larger for ST2, both models exhibit a significant change in curvature, reaching
or approaching a maximum in $U(\rho)$ near $\rho_{\rm eff}\approx0.80-0.90$.
The BKS model, on the other hand, has a minimum at $\rho_{\rm eff}\approx0.74$, a significantly larger value than for the other models.
While $U(\rho)$ for BKS does not approach a maximum, 
the range of data covers the appearance
of two inflection points.  
Once again, we see that the behavior of BKS is significantly different from that of the other models.

\subsection{Bond angle distributions}

\begin{figure}
\hbox to \hsize{\epsfxsize=1\hsize\epsfbox{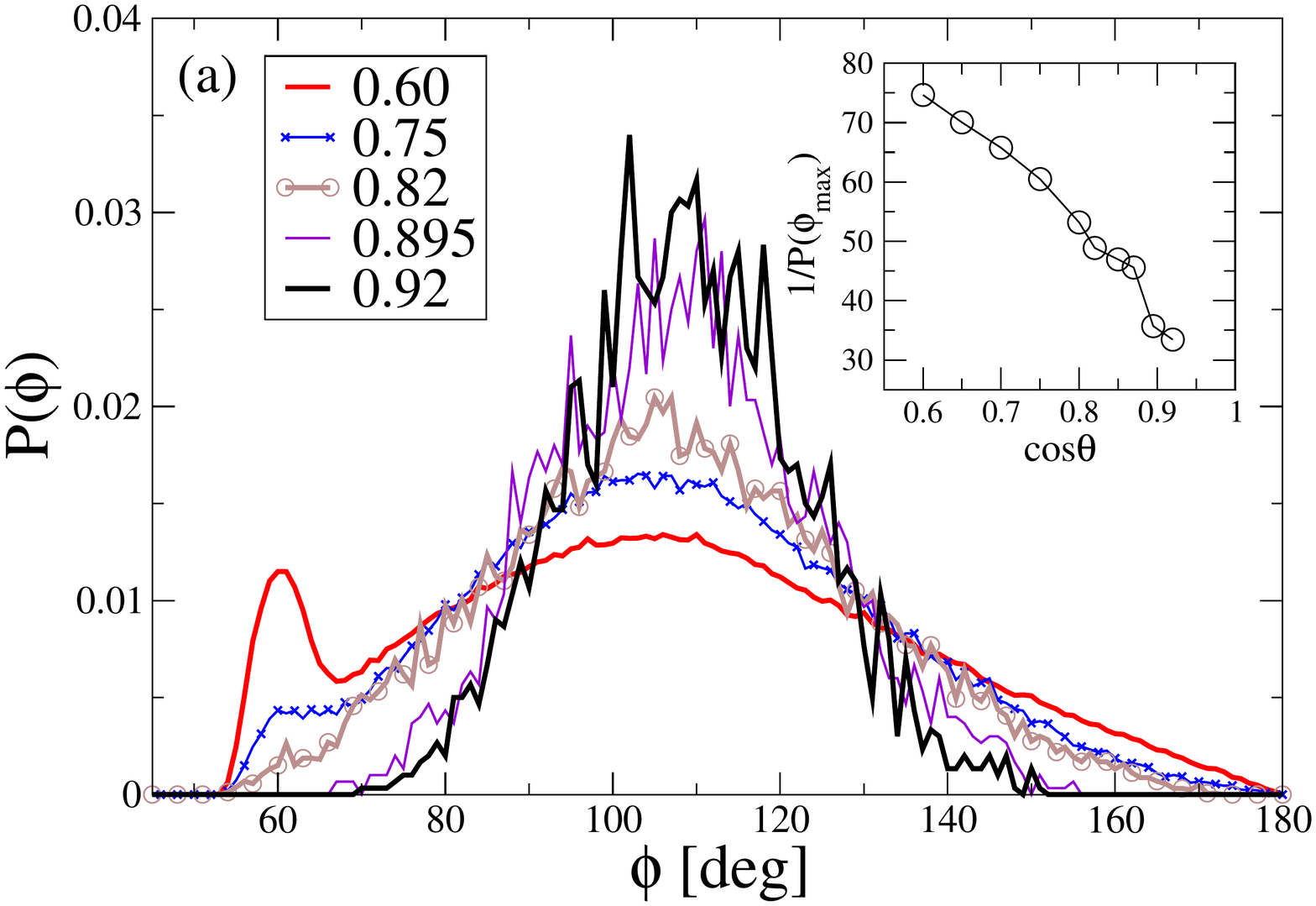}}
\hbox to \hsize{\epsfxsize=1\hsize\epsfbox{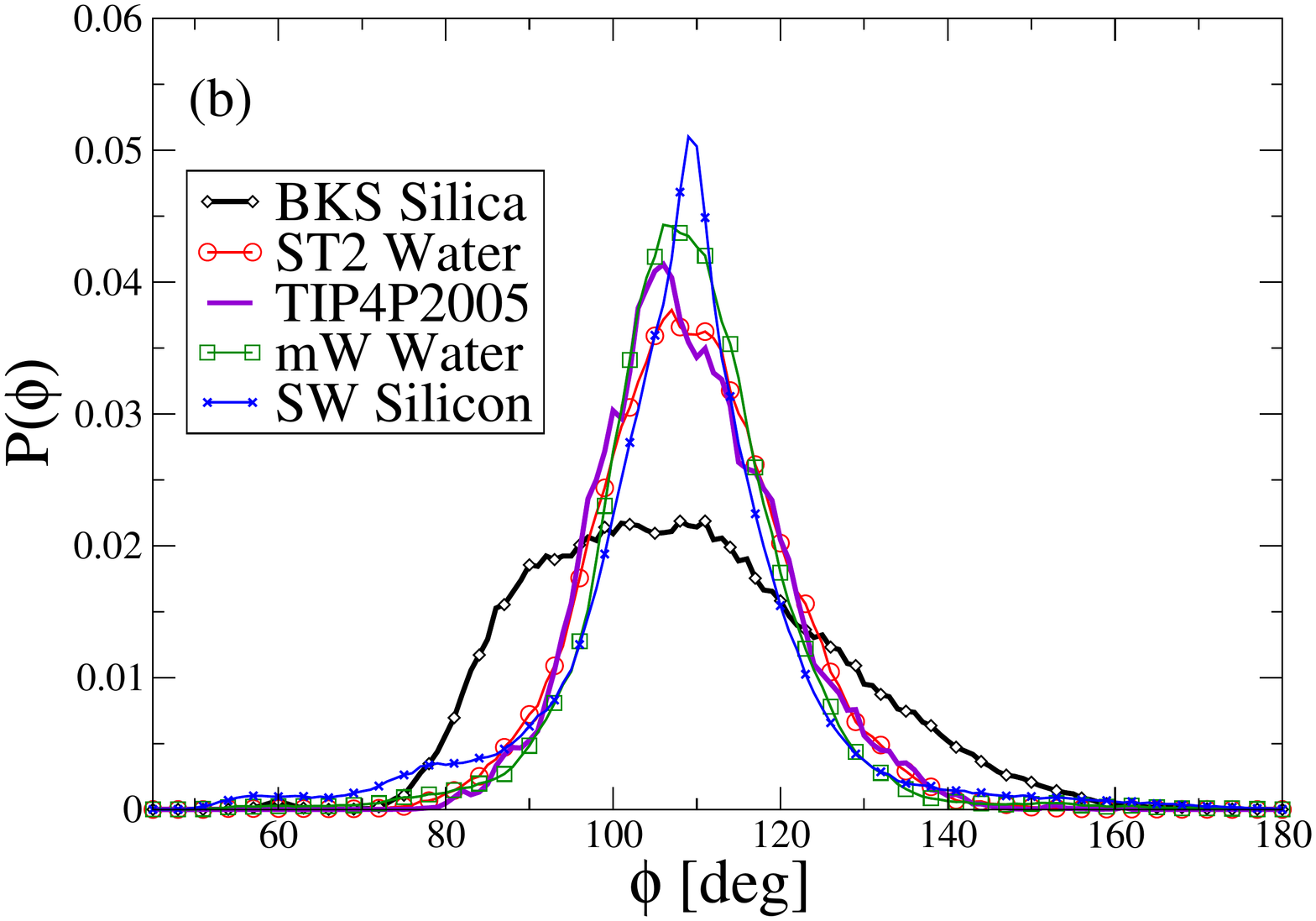}}
\caption{Probability distribution for the node-node-node bond angle.
Panel (a), KF model: As patches become narrower ($\cos\theta$ increases), 
there is a progressive narrowing of $P(\phi)$.  
Values of $\cos{\theta}$ are given in the legend.
Inset shows a measure of the width of the distribution, the inverse of the maximum probability $P(\phi_{\rm max})$,  as a function of $\cos\theta$.
Panel (b) shows $P(\phi)$ for the continuous potentials.
}
\label{figangle}
\end{figure}

\begin{figure}
\hbox to \hsize{\epsfxsize=1\hsize\epsfbox{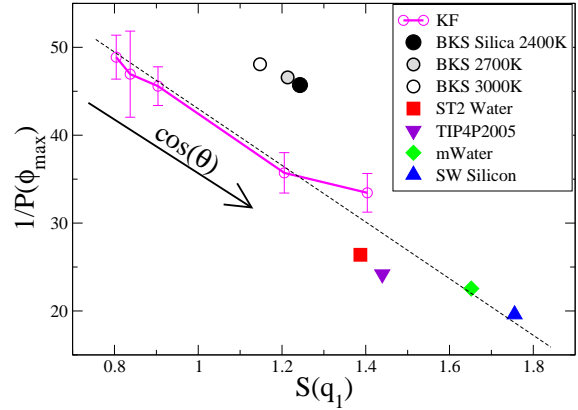}}
\caption{Inverse of the maximum probability $P(\phi_{\rm max})$ of the node-node-node angle distribution as a function of
$S(q_1)$, the height of the structure factor at the network peak.  The values of $\cos{\theta}$ for the KF models from left to right are
0.82, 0.85, 0.87, 0.895 and 0.92.  Dashed line is a guide to the eye.
}
\label{figp1invH}
\end{figure}

In Fig.~\ref{figangle}(a) we plot the probability density $P(\phi)$ for the node-node-node bond angle $\phi$ for the KF model
for a range of $\cos\theta$ from 0.60 to 0.92.  For wide patches, we see a significant peak at $\phi=60^\circ$ (arising from a significant
number of triangular rings) that disappears by $\cos\theta\approx 0.82$, coinciding with the appearance of the network peak in $S(q)$ 
and the appearance of a minimum in $U(\rho)$.  The main peak at the ideal tetrahedral angle of $109.5^\circ$ grows monotonically
with increasing $\cos\theta$ (narrowing patches), as the whole distribution narrows.  We define $P(\phi_{\rm max})$ as the maximum value
of $P(\phi)$, and take as a measure of the width of the distribution the inverse of this height, $1/P(\phi_{\rm max})$, 
partially to avoid difficulties associated with the peak at $60^\circ$.  This proxy for the width  is shown in the inset
to Fig.~\ref{figangle}(a) as a function of $\cos\theta$.  What is clear is that as the patch width narrows, so does the bond angle distribution.

We plot the distribution of angles for the various continuous models 
in Fig.~\ref{figangle}(b) and see 
that BKS has the broadest distribution.  BKS in fact looks to be the odd man out compared to the rest of the models, with a significant portion of its $P(\phi)$ deviating towards angles smaller than the ideal tetrahedral angle, about which the distributions
for the other models are peaked.  The tendency of silica to have a broader distribution than for water as well as a smaller average angle was already
noted in a previous study comparing primitive models of these network formers~\cite{DeMichele:2006iz}.
  
The peaks for the other models become progressively higher in the order of ST2, TIP4P/2005, mW and SW.  For mW and SW,  the kurtosis 
of the distribution is quite obviously positive.  This non-gaussian shape results in the standard deviation of the distributions not
becoming monotonically smaller as the peak height increases, and is another reason why we choose the inverse of peak height as a proxy for the
width.  Notwithstanding this detail, the heights of the distributions generally shadow the behavior of the height of the network peak in $S(q)$.

To make this point more clearly and motivated by the work Yuan and Cormack~\cite{Yuan:2003es} connecting bond angle distributions and particle 
correlations beyond first neighbors, 
we plot in Fig.~\ref{figp1invH} the quantity $1/P({\phi_{\rm max}})$ as a function of $S(q_1)$ for the KF
models with   $\cos\theta$ ranging over all the values for which we have a network peak, namely 0.82, 0.85, 0.87, 0.895 and 0.92, along 
with points corresponding to the continuous models.  With the exception of BKS silica, all the models fall near the same line.  
This linear relationship suggests that all the models (with the exception of silica) belong, in some sense, to the same family of tetrahedral models,
and that the members of this family with continuous potentials can be associated, even semiquantitatively, with a KF model with an appropriate 
patch width.  Clearly, the mW and SW models map onto KF models with much narrower patches than those to which ST2 and TIP4P/2005
map.  As for BKS, we can only qualitatively say that it would map onto a KF model with even wider patches.  

As a check on BKS silica, we plot in Fig.~\ref{figp1invH} data points for $T=3000$~K (open circle) and $T=2700$~K (grey circle) to get a sense
of how the values of $S(q_1)$ and $1/P({\phi_{\rm max}})$ vary over a small range in $T$.  We do not see a large variation in either quantity.
Neither do we see a trend that might suggest that for some lower $T$ the data would approach the data from the other models.

\subsection{Ring structure}

\begin{figure}
\hbox to \hsize{\epsfxsize=1\hsize\epsfbox{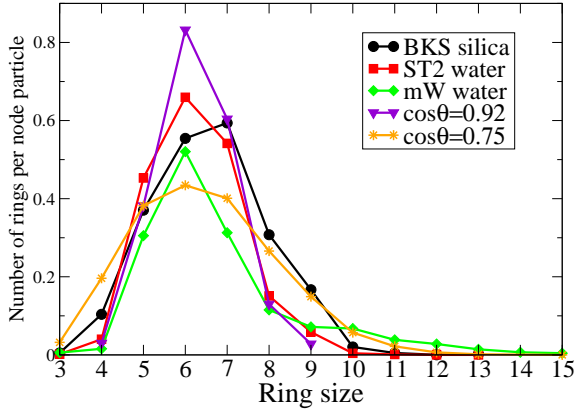}}
\caption{Ring size distribution for various models.
}
\label{figRings}
\end{figure}

To comment further on the uniqueness of BKS with respect to the other models, we plot in Fig.~\ref{figRings} the size distribution of minimal closed rings formed by neighboring node particles.  The KF model, with its tetrahedrally arranged patches, the ST2 model, wherein tetrahedrality
is encoded in the internal angle of the rigid water molecule and the mW model, in which a three-body term enforces a local tetrahedral 
geometry, all are peaked at a ring size of six, the number expected in crystals at this density.  BKS, on the other hand, has a peak at seven, and contains a significant number of rings of size eight and nine.  If BKS can be thought of as mapping onto a KF model with wide patches,
then it perhaps makes sense that the greater flexibility in node-node ``bonds'' allows for a broader distribution of ring sizes.  However, none of the wider patch KF models yield a similar ring structure, at least not at the density studied.  For example, shown in Fig.~\ref{figRings} is the ring size distribution for the $\cos\theta=0.75$ KF model, which has a similar number of rings of size eight and nine compared to BKS.  However, as is plainly evident, the rest of the curve is quite different.

\subsection{KF as ST2 and BKS}

\begin{figure}
\hbox to \hsize{\epsfxsize=1\hsize\epsfbox{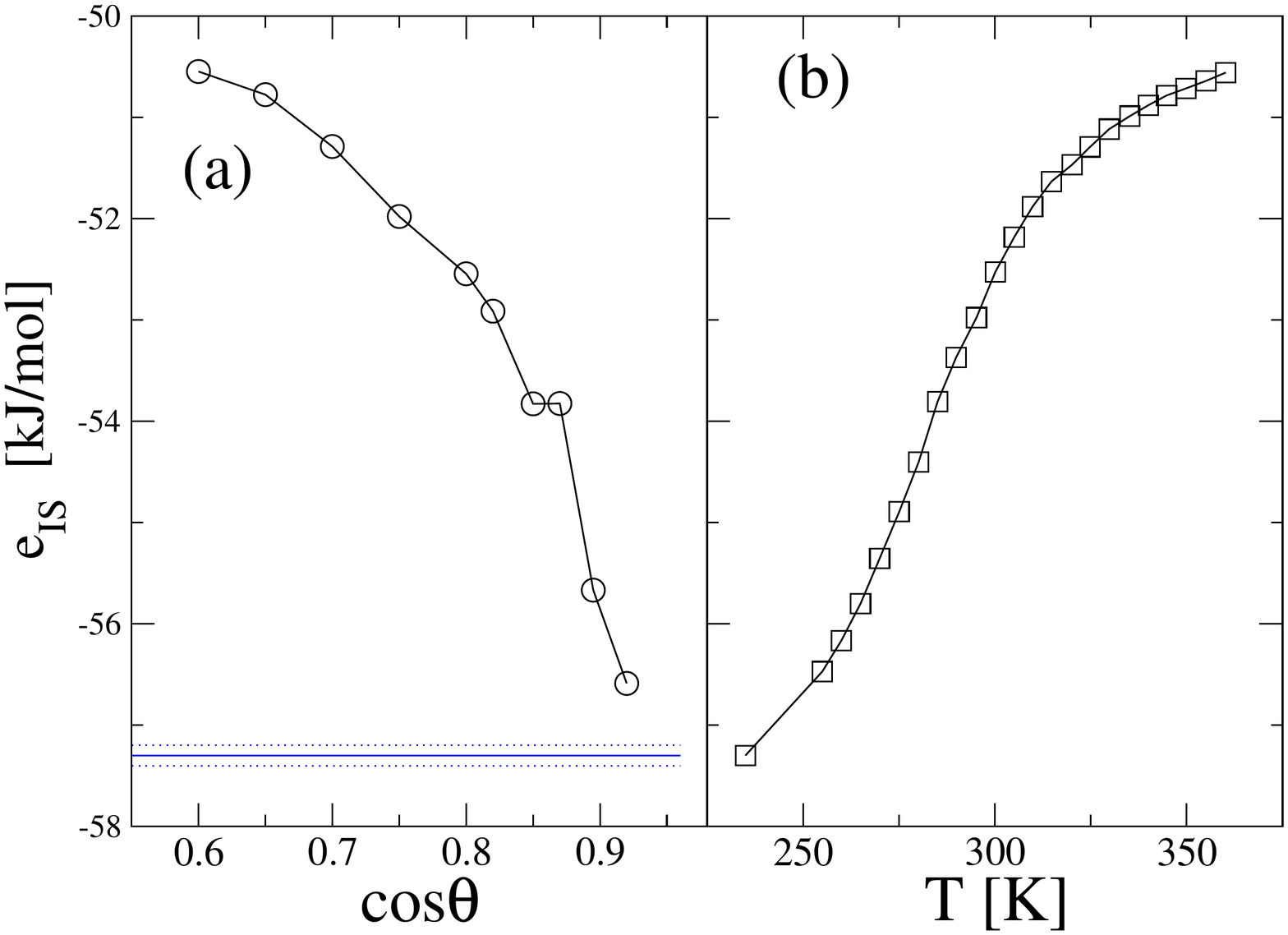}}
\hbox to \hsize{\epsfxsize=1\hsize\epsfbox{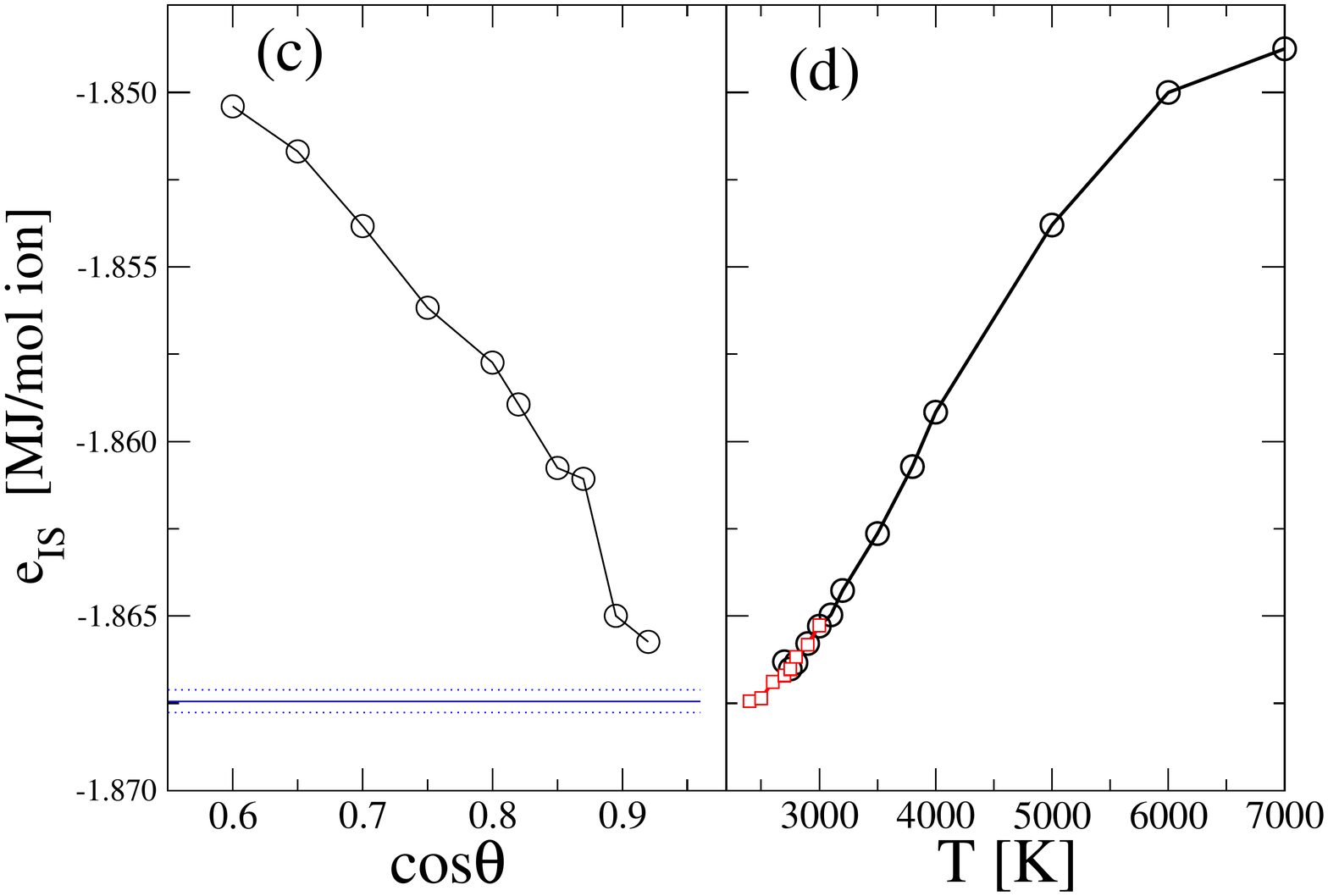}}
\caption{
Energy of inherent structures, $e_{\rm IS}$, for (a) ST2 at 0.83 g/cm$^3$ and 
(c) BKS at 2.36 g/cm$^3$ generated from fully bonded KF configurations at $\rho \sigma^3=0.57$ 
with varying patch width.  Horizontal lines
indicate the  $e_{\rm IS}$ obtained from direct simulations of the original models at the lowest $T$ probed and same density, 
with dotted lines indicating standard deviation.
Also shown are the full $e_{\rm IS}(T)$ curves obtained from simulations of (b) ST2 
(the point at $T=235$~K is obtained in this work, the points at
higher $T$ points are taken from Ref.~\cite{Poole:2011bb}) and (d) BKS
(squares indicate data obtained in this work, circles from Ref.~\cite{SaikaVoivod:2004jf}).
}
\label{figenergy}
\end{figure}

The fact that the KF model in the singe-particle-per-patch regime, whether enforced or not by the special MD procedure, forms
networks with all possible bonds satisfied, allows for the intriguing possibility of using KF configurations as starting points for 
searching for the lowest energy amorphous configurations in other models.  To this end, we study the IS energy of KF configurations
after they are converted into ST2 water at $\rho=0.83$~g/cm$^3$ and BKS silica at $\rho=2.36$~g/cm$^3$ .

For ST2, since the rigid molecule already has the same HOH angle as the angle that separates patches on the KF particles, the main
complication in converting a fully bonded KF configuration to ST2 is appropriately ``filling'' only two of four patches of each particle with protons.
To do so, we randomly choose a particle that has fewer than two protons assigned and randomly fill with a proton an unfilled patch that is not
overlapping with a filled patch from a different particle.
This chosen patch is necessarily bonded to another particle in need of at least one proton, and so one of
the unfilled patches of this second particle, excluding any patches overlapping with a filled patch 
(such as the one connecting it back to the first particle), 
is selected at random to fill.  
This is repeated until the procedure loops back to the initial particle.  The net dipole moment of the loop is nearly zero.  The procedure is iterated
by randomly selecting a particle to start a new loop
until all particles have been visited twice.  This method of constructing proton disorder avoids artifacts due to a significant net dipole moment in the
simulation cell~\cite{dip1,dip2}.


After the conversion is made, 
a CG quench is performed to obtain $e_{\rm IS}$.  
The results are plotted in Fig.~\ref{figenergy}(a),
and show that while at low values of $\cos\theta$, $e_{\rm IS}$ is rather high, there is a steep drop, with an extrapolated intersection
with the average $e_{\rm IS}$ obtained from simulations of ST2 at $T=235$~K, the lowest $e_{\rm IS}$ that we have, at about 
$\cos\theta=0.94$.  For comparison, in Fig.~\ref{figenergy}(b) we show $e_{\rm IS}(T)$ for ST2, with points above $T=250$~K taken from
Ref.~\cite{Poole:2011bb}.  The KF configurations can generally produce ST2 configurations that are quite low in the potential energy landscape.

In order to convert KF into BKS, we take the KF particle positions as the positions of the Si ions and then place
O ions at the midway points between bonded KF particles, scaling all coordinates as required.  We then perform a CG quench on the system.
We have checked other procedures for placing the O ions, for example, placing them midway between the centres of overlapping patches,
and have tried quenching only O ion positions first, and then quenching the entire system, but the gain in energy is somewhat marginal.
The results of the conversion are plotted in Fig.~\ref{figenergy}(c).  While we do not have sufficient numbers of perfect KF networks
at large values of $\cos\theta$ to make very precise statements, 
it does not seem improbable to us that the data from lower values of $\cos\theta$ suggest an extrapolated
intersection with our best $e_{\rm IS}$ for BKS (from $T=2400$~K) at a value of $\cos\theta$ lower than that apparent for ST2, but then
rather than achieving the best BKS $e_{\rm IS}$, the $e_{\rm IS}(\cos\theta)$ curve inflects away from the $T=2400$~K $e_{\rm IS}$ line.
Such an interpretation would be consistent with the results in the previous sections that suggest BKS corresponds to a KF model with wider patches than ST2.

For comparison, we plot $e_{\rm IS}(T)$ for BKS in Fig.~\ref{figenergy}(d).  Two curves are present, one taken from Ref.~\cite{SaikaVoivod:2004jf}, the 
other, at lower $T$, obtained for the present study.   For the new data, we remind the reader that we used the original version of BKS augmented 
by a potential at very short ranges to prevent ``fusion'' events and then quenched using the potential used in Ref.~\cite{SaikaVoivod:2004jf}.  That the
curves coincide in the region of overlap shows that the additional modifications to BKS in Ref.~\cite{SaikaVoivod:2004jf}, namely a fixed Ewald parameter
and a tapering of the real space potential energy from 0.77476~nm to zero at 1~nm, do not produce significantly different inherent structures
from the original BKS potential.

\section{Discussion}

The KF model, augmented with a constraint that ensures only one bond per patch, 
allows one to unambiguously define a fully bonded energetic ground state even for very wide patches.
We study the properties of the ground state of the tetrahedral version of the model as a function of patch width and find
that as the patches become narrower, the system becomes progressively more ordered.  We mean this in the specific sense that
by $\cos\theta\approx0.8$, the system develops a peak in $S(q)$ characteristic of network-forming liquids, a minimum in the 
potential energy as a function of density corresponding to an optimal network-forming density and a disappearance of the secondary
peak in the bond angle distribution at $60^\circ$.
The bond angle distribution becomes progressively more peaked as patches become narrower and there is a linear relationship between 
the width of the angle distribution and the height of the network peak in $S(q)$.  

By working at optimal network-forming densities 
with other, continuous models of network-forming liquids such as silica, water and silicon, we remove density as a parameter
in an endeavor  to roughly map the behavior of these models to the KF model with patch width as the single mapping parameter.
As Fig.~\ref{figp1invH} shows, with the exception of BKS silica, all models studied fall very near the same behavior, and it is thus possible,
in principle, to identify the various continuous models with effective KF models of varying patch widths.
This is particularly useful in light of recent studies of the KF model pertaining to crystallization~\cite{Romano:2011jm,SaikaVoivod:2011co} 
and liquid stability~\cite{Smallenburg:2013cq}

The most tetrahedrally constrained of the models, namely mW water and SW silicon, correspond to narrow patches.  The three-body 
potential in these models confer a lack of bond flexibility, as do narrow patches in the KF model.   The mW and SW models 
share an important characteristic of the KF models with large $\cos\theta$, namely the difficulty of reaching the energetic ground state before 
crystallizing.  The networks of mW and SW, at the $T$ we use, are quite imperfect compared to those of BKS, ST2 and TIP4P/2005,
and accessing progressively better networks by lowering $T$ is difficult because of crystallization.

Tetrahedrality in ST2 and TIP4P/2005 is less stringently enforced, and arises from hydrogen bonding between tetrahedrally bent rigid
water molecules.  Thus, there is more inherent flexibility in the node-node bonds.  This observation is consistent with ST2 and TIP4P/2005
mapping to values of $\cos\theta$ in the range 0.92 to 0.94, and the fact that nearly defect-free networks within these models are attainable
without crystallization.

The network of BKS, in light of the present results, is significantly non-tetrahedral.  Tetrahedrality in BKS is even less strictly enforced,
provided essentially by stoichiometry, steric repulsion and charge.  The node-node ``bond'' is mediated by a relatively large O ion,
allowing for a great deal a flexibility.  The structural quantities of BKS we study, namely, $S(q)$, $P(\phi)$ and ring size distributions
are qualitatively different from the other models.  Nonetheless, perhaps in a more qualitative way, insights into the the properties of 
the BKS network can be gained from understanding the properties of KF models with wide patches.

In Ref.~\cite{Romano:2011jm}, the authors showed that the driving force for nucleation, i.e., the chemical potential difference between 
crystal and liquid $\Delta\mu$ becomes progressively smaller as $\cos\theta$ decreases, and a subsequent study of the
nucleation barriers enforced the idea that for $\cos\theta\approx 0.92$ and below, nucleation becomes difficult and the system 
essentially becomes a glass former.  This idea was carried to the extreme in Ref.~\cite{Smallenburg:2013cq}, where the authors showed that
if the one-bond-per-patch condition is maintained, then the liquid remains as the stable phase down to $T=0$ at the expense 
of the BCC crystal for a range
of densities, certainly for values of $\cos\theta$ less than or equal to 0.80.  This value of 0.80 is interesting in that it also represents
the cusp of the KF liquid becoming a networked liquid, rather than simply being four-coordinated.

This idea of increasing patch width to increase liquid stability with respect to crystallization is consistent with previous work done on the
SW family of potentials wherein the three-body constraint strength $\lambda$ was tuned~\cite{Molinero:2006hk}.  At weaker $\lambda$, the liquid was
stable to progressively lower $T$.  Were it not for the appearance of the BCC crystal, perhaps the $T=0$ limit could be reached.
\\


Obtaining perfect networks for narrow patches is inherently difficult.  The rigid geometry enforced by narrow patches or inflexible
bonds implies that low energy configurations must resemble crystals, or at least have a reduced number of ways in which a random network
can be formed, i.e., a reduction in the configurational entropy and therefore a reduction in liquid stability.  If one wished to map the
KF model to models such as mW and SW in a more precise way in spite of this difficulty, 
one could perhaps compare the state points with similar defect concentration.


\section{Conclusions}

We study a family of fully bonded tetrahedral KF patchy particle systems as a function of patch width at the optimal network-forming
density and find a few concurrent measures for the onset of a structured network at $\cos{\theta}\approx0.80$.  
There is a linear relationship between the width of the bond angle distribution and the height of the network peak in the 
structure factor.  Several other models follow this ``family line'', including the ST2, TIP4P/2005 and mW models of water, as well as
SW silicon.   This suggests a mapping of these models to KF models of different patch widths.  This mapping makes intuitive
sense given the nature of the potentials and degree of bond flexibility in each of the models.

The mapping is useful given the systematic study of how the patch width affects the ability of the KF model to crystallize, or
conversely, to avoid crystallization.  Essentially, the narrower the patches, the greater the propensity for the model to crystallize.
Wider  patches allow for the system to approach or even reach the ground state energy and avoid nucleation~\cite{Romano:2010bl,Romano:2011jm, SaikaVoivod:2011co,Smallenburg:2013cq}.
According to the 
semi-qualitative mapping, SW and mW map to quite narrow patches, and this explains why they are prone to crystallize before
achieving a relatively defect-free network.  ST2 and TIP4P/2005, on the other hand, map to wider patches and are therefore  better network glass-formers.   The perfect correspondence between
crystallization ability and bonding angular width observed throughout all these models reinforces the
general validity of the results obtained in the investigation of the KF model.

BKS silica, the best glass former,  is qualitatively different in terms of several properties studied and does not fall near the family line.  Its network properties are rather different. Possibly, one needs to devise a binary mixture analog~\cite{monson,DeMichele:2006iz,bianchisilica,SaikaVoivod:2011by}  in order to capture the essential differences.
A form of the KF model has also been recently shown to have a liquid that is the thermodynamic ground state for a range of densities for sufficiently wide patches~\cite{Smallenburg:2013cq}.  It remains a challenge to alter the continuous molecular potentials in order to achieve such a liquid ground
state in another class of potentials. 

Finally, we like to note that man-made particles can help shed light on unsolved problems
in atomic and molecular physics, in the present case connecting  gel-forming patchy colloids and 
tetrahedral network glass formers~\cite{statphys}. New soft-matter systems in which valence can be precisely controlled, e.g., DNA constructs~\cite{luoxy,biffi}  and new polymers~\cite{vitrimers}, may contribute to
deepening our understanding of fundamental problems in disordered systems.
 Providing valence to colloids~\cite{mohovald,Wang:2012gd}  is opening
a very rich line of investigation.

\section*{Acknowledgments}

IS-V thanks NSERC and ERC-PATCHYCOLLOIDS for funding, ACEnet for computational support, CFI for funding of computing infrastructure, and Sapienza University for hosting.  Both F.S. acknowledge
support from  ERC-226207-PATCHYCOLLOIDS and MIUR-PRIN.


\begin{thebibliography}{61}
\expandafter\ifx\csname natexlab\endcsname\relax\def\natexlab#1{#1}\fi
\expandafter\ifx\csname bibnamefont\endcsname\relax
  \def\bibnamefont#1{#1}\fi
\expandafter\ifx\csname bibfnamefont\endcsname\relax
  \def\bibfnamefont#1{#1}\fi
\expandafter\ifx\csname citenamefont\endcsname\relax
  \def\citenamefont#1{#1}\fi
\expandafter\ifx\csname url\endcsname\relax
  \def\url#1{\texttt{#1}}\fi
\expandafter\ifx\csname urlprefix\endcsname\relax\def\urlprefix{URL }\fi
\providecommand{\bibinfo}[2]{#2}
\providecommand{\eprint}[2][]{\url{#2}}

\bibitem[{\citenamefont{Binder and Kob}(2005)}]{kob-binder-book}
\bibinfo{author}{\bibfnamefont{K.}~\bibnamefont{Binder}} \bibnamefont{and}
  \bibinfo{author}{\bibfnamefont{W.}~\bibnamefont{Kob}},
  \emph{\bibinfo{title}{Glassy Materials And Disordered Solids: An Introduction
  to Their Statistical Mechanics}} (\bibinfo{publisher}{World Scientific
  Publishing Company}, \bibinfo{year}{2005}), ISBN
  \bibinfo{isbn}{9789812565105}.

\bibitem[{\citenamefont{Pawar and Kretzschmar}(2010)}]{Pawar:2010ig}
\bibinfo{author}{\bibfnamefont{A.~B.} \bibnamefont{Pawar}} \bibnamefont{and}
  \bibinfo{author}{\bibfnamefont{I.}~\bibnamefont{Kretzschmar}},
  \bibinfo{journal}{Macromol. Rapid Commun.} \textbf{\bibinfo{volume}{31}},
  \bibinfo{pages}{150} (\bibinfo{year}{2010}).

\bibitem[{\citenamefont{Bianchi et~al.}(2011)\citenamefont{Bianchi, Blaak, and
  Likos}}]{BianchiPCCP}
\bibinfo{author}{\bibfnamefont{E.}~\bibnamefont{Bianchi}},
  \bibinfo{author}{\bibfnamefont{R.}~\bibnamefont{Blaak}}, \bibnamefont{and}
  \bibinfo{author}{\bibfnamefont{C.~N.} \bibnamefont{Likos}},
  \bibinfo{journal}{Phys. Chem.} \textbf{\bibinfo{volume}{13}},
  \bibinfo{pages}{6397} (\bibinfo{year}{2011}).

\bibitem[{\citenamefont{Wang et~al.}(2012)\citenamefont{Wang, Wang, Breed,
  Manoharan, Feng, Hollingsworth, Weck, and Pine}}]{Wang:2012gd}
\bibinfo{author}{\bibfnamefont{Y.}~\bibnamefont{Wang}},
  \bibinfo{author}{\bibfnamefont{Y.}~\bibnamefont{Wang}},
  \bibinfo{author}{\bibfnamefont{D.~R.} \bibnamefont{Breed}},
  \bibinfo{author}{\bibfnamefont{V.~N.} \bibnamefont{Manoharan}},
  \bibinfo{author}{\bibfnamefont{L.}~\bibnamefont{Feng}},
  \bibinfo{author}{\bibfnamefont{A.~D.} \bibnamefont{Hollingsworth}},
  \bibinfo{author}{\bibfnamefont{M.}~\bibnamefont{Weck}}, \bibnamefont{and}
  \bibinfo{author}{\bibfnamefont{D.~J.} \bibnamefont{Pine}},
  \bibinfo{journal}{Nature} \textbf{\bibinfo{volume}{490}}, \bibinfo{pages}{51}
  (\bibinfo{year}{2012}).

\bibitem[{\citenamefont{Biffi et~al.}(2013)\citenamefont{Biffi, Cerbino,
  Bomboi, Paraboschi, Asselta, Sciortino, and Bellini}}]{biffi}
\bibinfo{author}{\bibfnamefont{S.}~\bibnamefont{Biffi}},
  \bibinfo{author}{\bibfnamefont{R.}~\bibnamefont{Cerbino}},
  \bibinfo{author}{\bibfnamefont{F.}~\bibnamefont{Bomboi}},
  \bibinfo{author}{\bibfnamefont{E.~M.} \bibnamefont{Paraboschi}},
  \bibinfo{author}{\bibfnamefont{R.}~\bibnamefont{Asselta}},
  \bibinfo{author}{\bibfnamefont{F.}~\bibnamefont{Sciortino}},
  \bibnamefont{and} \bibinfo{author}{\bibfnamefont{T.}~\bibnamefont{Bellini}},
  \bibinfo{journal}{PNAS (in press)}  (\bibinfo{year}{2013}).

\bibitem[{\citenamefont{Chen et~al.}(2011)\citenamefont{Chen, Bae, and
  Granick}}]{Granick}
\bibinfo{author}{\bibfnamefont{Q.}~\bibnamefont{Chen}},
  \bibinfo{author}{\bibfnamefont{S.~C.} \bibnamefont{Bae}}, \bibnamefont{and}
  \bibinfo{author}{\bibfnamefont{S.}~\bibnamefont{Granick}},
  \bibinfo{journal}{Nature} \textbf{\bibinfo{volume}{469}},
  \bibinfo{pages}{381} (\bibinfo{year}{2011}).

\bibitem[{\citenamefont{Vlcek and Nezbeda}(2004)}]{Vlc04a}
\bibinfo{author}{\bibfnamefont{L.}~\bibnamefont{Vlcek}} \bibnamefont{and}
  \bibinfo{author}{\bibfnamefont{I.}~\bibnamefont{Nezbeda}},
  \bibinfo{journal}{Mol. Phys.} \textbf{\bibinfo{volume}{102}},
  \bibinfo{pages}{771} (\bibinfo{year}{2004}).

\bibitem[{\citenamefont{Nezbeda}(2005)}]{Nez05a}
\bibinfo{author}{\bibfnamefont{I.}~\bibnamefont{Nezbeda}},
  \bibinfo{journal}{Mol. Phys.} \textbf{\bibinfo{volume}{103}},
  \bibinfo{pages}{59} (\bibinfo{year}{2005}).

\bibitem[{\citenamefont{Kolafa and Nezbeda}(1987)}]{Kolafa:1987wk}
\bibinfo{author}{\bibfnamefont{J.}~\bibnamefont{Kolafa}} \bibnamefont{and}
  \bibinfo{author}{\bibfnamefont{I.}~\bibnamefont{Nezbeda}},
  \bibinfo{journal}{Mol. Phys.} \textbf{\bibinfo{volume}{61}},
  \bibinfo{pages}{161} (\bibinfo{year}{1987}).

\bibitem[{\citenamefont{Kern and Frenkel}(2003)}]{Kern:2003ie}
\bibinfo{author}{\bibfnamefont{N.}~\bibnamefont{Kern}} \bibnamefont{and}
  \bibinfo{author}{\bibfnamefont{D.}~\bibnamefont{Frenkel}},
  \bibinfo{journal}{J. Chem. Phys.} \textbf{\bibinfo{volume}{118}},
  \bibinfo{pages}{9882} (\bibinfo{year}{2003}).

\bibitem[{\citenamefont{Bol}(1982)}]{bol}
\bibinfo{author}{\bibfnamefont{W.}~\bibnamefont{Bol}}, \bibinfo{journal}{Mol.
  Phys.} \textbf{\bibinfo{volume}{45}}, \bibinfo{pages}{605}
  (\bibinfo{year}{1982}).

\bibitem[{\citenamefont{Romano et~al.}(2010)\citenamefont{Romano, Sanz, and
  Sciortino}}]{Romano:2010bl}
\bibinfo{author}{\bibfnamefont{F.}~\bibnamefont{Romano}},
  \bibinfo{author}{\bibfnamefont{E.}~\bibnamefont{Sanz}}, \bibnamefont{and}
  \bibinfo{author}{\bibfnamefont{F.}~\bibnamefont{Sciortino}},
  \bibinfo{journal}{J. Chem. Phys.} \textbf{\bibinfo{volume}{132}},
  \bibinfo{pages}{184501} (\bibinfo{year}{2010}).

\bibitem[{\citenamefont{Romano et~al.}(2011)\citenamefont{Romano, Sanz, and
  Sciortino}}]{Romano:2011jm}
\bibinfo{author}{\bibfnamefont{F.}~\bibnamefont{Romano}},
  \bibinfo{author}{\bibfnamefont{E.}~\bibnamefont{Sanz}}, \bibnamefont{and}
  \bibinfo{author}{\bibfnamefont{F.}~\bibnamefont{Sciortino}},
  \bibinfo{journal}{J. Chem. Phys.} \textbf{\bibinfo{volume}{134}},
  \bibinfo{pages}{174502} (\bibinfo{year}{2011}).

\bibitem[{\citenamefont{Saika-Voivod
  et~al.}(2011{\natexlab{a}})\citenamefont{Saika-Voivod, Romano, and
  Sciortino}}]{SaikaVoivod:2011co}
\bibinfo{author}{\bibfnamefont{I.}~\bibnamefont{Saika-Voivod}},
  \bibinfo{author}{\bibfnamefont{F.}~\bibnamefont{Romano}}, \bibnamefont{and}
  \bibinfo{author}{\bibfnamefont{F.}~\bibnamefont{Sciortino}},
  \bibinfo{journal}{J. Chem. Phys.} \textbf{\bibinfo{volume}{135}},
  \bibinfo{pages}{124506} (\bibinfo{year}{2011}{\natexlab{a}}).

\bibitem[{\citenamefont{Smallenburg and Sciortino}(2013)}]{Smallenburg:2013cq}
\bibinfo{author}{\bibfnamefont{F.}~\bibnamefont{Smallenburg}} \bibnamefont{and}
  \bibinfo{author}{\bibfnamefont{F.}~\bibnamefont{Sciortino}},
  \bibinfo{journal}{Nat. Phys.} \textbf{\bibinfo{volume}{9}},
  \bibinfo{pages}{554–} (\bibinfo{year}{2013}).

\bibitem[{\citenamefont{van Beest et~al.}(1990)\citenamefont{van Beest, Kramer,
  and van Santen}}]{vanBeest:1990tt}
\bibinfo{author}{\bibfnamefont{B.~W.~H.} \bibnamefont{van Beest}},
  \bibinfo{author}{\bibfnamefont{G.~J.} \bibnamefont{Kramer}},
  \bibnamefont{and} \bibinfo{author}{\bibfnamefont{R.~A.} \bibnamefont{van
  Santen}}, \bibinfo{journal}{Phys. Rev. Lett.} \textbf{\bibinfo{volume}{64}},
  \bibinfo{pages}{1995} (\bibinfo{year}{1990}).

\bibitem[{\citenamefont{Stillinger and Rahman}(1974)}]{Stillinger:1974fh}
\bibinfo{author}{\bibfnamefont{F.~H.} \bibnamefont{Stillinger}}
  \bibnamefont{and} \bibinfo{author}{\bibfnamefont{A.}~\bibnamefont{Rahman}},
  \bibinfo{journal}{J. Chem. Phys.} \textbf{\bibinfo{volume}{60}},
  \bibinfo{pages}{1545} (\bibinfo{year}{1974}).

\bibitem[{\citenamefont{Abascal and Vega}(2005)}]{Abascal:2005ka}
\bibinfo{author}{\bibfnamefont{J.~L.~F.} \bibnamefont{Abascal}}
  \bibnamefont{and} \bibinfo{author}{\bibfnamefont{C.}~\bibnamefont{Vega}},
  \bibinfo{journal}{J. Chem. Phys.} \textbf{\bibinfo{volume}{123}},
  \bibinfo{pages}{234505} (\bibinfo{year}{2005}).

\bibitem[{\citenamefont{Molinero and Moore}(2009)}]{Molinero:2009kc}
\bibinfo{author}{\bibfnamefont{V.}~\bibnamefont{Molinero}} \bibnamefont{and}
  \bibinfo{author}{\bibfnamefont{E.~B.} \bibnamefont{Moore}},
  \bibinfo{journal}{J. Phys. Chem. B} \textbf{\bibinfo{volume}{113}},
  \bibinfo{pages}{4008} (\bibinfo{year}{2009}).

\bibitem[{\citenamefont{Stillinger and Weber}(1985)}]{Stillinger:1985vx}
\bibinfo{author}{\bibfnamefont{F.~H.} \bibnamefont{Stillinger}}
  \bibnamefont{and} \bibinfo{author}{\bibfnamefont{T.~A.} \bibnamefont{Weber}},
  \bibinfo{journal}{Phys. Rev. B} \textbf{\bibinfo{volume}{31}},
  \bibinfo{pages}{5262} (\bibinfo{year}{1985}).

\bibitem[{\citenamefont{De~Michele
  et~al.}(2006{\natexlab{a}})\citenamefont{De~Michele, Gabrielli, Tartaglia,
  and Sciortino}}]{simone}
\bibinfo{author}{\bibfnamefont{C.}~\bibnamefont{De~Michele}},
  \bibinfo{author}{\bibfnamefont{S.}~\bibnamefont{Gabrielli}},
  \bibinfo{author}{\bibfnamefont{P.}~\bibnamefont{Tartaglia}},
  \bibnamefont{and}
  \bibinfo{author}{\bibfnamefont{F.}~\bibnamefont{Sciortino}},
  \bibinfo{journal}{J. Phys. Chem. B} \textbf{\bibinfo{volume}{110}},
  \bibinfo{pages}{8064} (\bibinfo{year}{2006}{\natexlab{a}}).

\bibitem[{\citenamefont{Zachariasen}(1932)}]{Zachariasen:1932tn}
\bibinfo{author}{\bibfnamefont{W.~H.} \bibnamefont{Zachariasen}},
  \bibinfo{journal}{J. Amer. Chem. Soc.} \textbf{\bibinfo{volume}{54}},
  \bibinfo{pages}{3841} (\bibinfo{year}{1932}).

\bibitem[{\citenamefont{Evans and King}(1966)}]{Evans:1966vk}
\bibinfo{author}{\bibfnamefont{D.~L.} \bibnamefont{Evans}} \bibnamefont{and}
  \bibinfo{author}{\bibfnamefont{S.~V.} \bibnamefont{King}},
  \bibinfo{journal}{Nature} \textbf{\bibinfo{volume}{212}},
  \bibinfo{pages}{1353} (\bibinfo{year}{1966}).

\bibitem[{\citenamefont{Bell and Dean}(1966)}]{Bell:1966ww}
\bibinfo{author}{\bibfnamefont{R.~J.} \bibnamefont{Bell}} \bibnamefont{and}
  \bibinfo{author}{\bibfnamefont{P.}~\bibnamefont{Dean}},
  \bibinfo{journal}{Nature} \textbf{\bibinfo{volume}{212}},
  \bibinfo{pages}{1354} (\bibinfo{year}{1966}).

\bibitem[{\citenamefont{Sceats et~al.}(1979)\citenamefont{Sceats, Stavola, and
  Rice}}]{Sceats:1979em}
\bibinfo{author}{\bibfnamefont{M.~G.} \bibnamefont{Sceats}},
  \bibinfo{author}{\bibfnamefont{M.}~\bibnamefont{Stavola}}, \bibnamefont{and}
  \bibinfo{author}{\bibfnamefont{S.~A.} \bibnamefont{Rice}},
  \bibinfo{journal}{J. Chem. Phys.} \textbf{\bibinfo{volume}{70}},
  \bibinfo{pages}{3927} (\bibinfo{year}{1979}).

\bibitem[{\citenamefont{Rice and Sceats}(1981)}]{Rice:1981vz}
\bibinfo{author}{\bibfnamefont{S.~A.} \bibnamefont{Rice}} \bibnamefont{and}
  \bibinfo{author}{\bibfnamefont{M.~G.} \bibnamefont{Sceats}},
  \bibinfo{journal}{J. Chem. Phys.} \textbf{\bibinfo{volume}{85}},
  \bibinfo{pages}{1108} (\bibinfo{year}{1981}).

\bibitem[{\citenamefont{Henn and Kauzmann}(1989)}]{Henn:1989ua}
\bibinfo{author}{\bibfnamefont{A.~R.} \bibnamefont{Henn}} \bibnamefont{and}
  \bibinfo{author}{\bibfnamefont{W.}~\bibnamefont{Kauzmann}},
  \bibinfo{journal}{J. Phys. C: Solid State Phys.}
  \textbf{\bibinfo{volume}{93}}, \bibinfo{pages}{3770} (\bibinfo{year}{1989}).

\bibitem[{\citenamefont{Polk and Boudreaux}(1973)}]{Polk:73wr}
\bibinfo{author}{\bibfnamefont{D.~E.} \bibnamefont{Polk}} \bibnamefont{and}
  \bibinfo{author}{\bibfnamefont{D.~S.} \bibnamefont{Boudreaux}},
  \bibinfo{journal}{Phys. Rev. Lett.} \textbf{\bibinfo{volume}{31}},
  \bibinfo{pages}{92} (\bibinfo{year}{1973}).

\bibitem[{\citenamefont{Molinero et~al.}(2006)\citenamefont{Molinero, Sastry,
  and Angell}}]{Molinero:2006hk}
\bibinfo{author}{\bibfnamefont{V.}~\bibnamefont{Molinero}},
  \bibinfo{author}{\bibfnamefont{S.}~\bibnamefont{Sastry}}, \bibnamefont{and}
  \bibinfo{author}{\bibfnamefont{C.~A.} \bibnamefont{Angell}},
  \bibinfo{journal}{Phys. Rev. Lett.} \textbf{\bibinfo{volume}{97}},
  \bibinfo{pages}{075701} (\bibinfo{year}{2006}).

\bibitem[{\citenamefont{Wooten et~al.}(1985)\citenamefont{Wooten, Winer, and
  Weaire}}]{Wooten:1985vx}
\bibinfo{author}{\bibfnamefont{F.}~\bibnamefont{Wooten}},
  \bibinfo{author}{\bibfnamefont{K.}~\bibnamefont{Winer}}, \bibnamefont{and}
  \bibinfo{author}{\bibfnamefont{D.}~\bibnamefont{Weaire}},
  \bibinfo{journal}{Phys. Rev. Lett.} \textbf{\bibinfo{volume}{54}},
  \bibinfo{pages}{1392} (\bibinfo{year}{1985}).

\bibitem[{\citenamefont{Barkema and Mousseau}(2000)}]{Barkema:2000ut}
\bibinfo{author}{\bibfnamefont{G.~T.} \bibnamefont{Barkema}} \bibnamefont{and}
  \bibinfo{author}{\bibfnamefont{N.}~\bibnamefont{Mousseau}},
  \bibinfo{journal}{Phys. Rev. B} \textbf{\bibinfo{volume}{62}},
  \bibinfo{pages}{4985} (\bibinfo{year}{2000}).

\bibitem[{\citenamefont{Mousseau and Barkema}(2004)}]{Mousseau:2004fk}
\bibinfo{author}{\bibfnamefont{N.}~\bibnamefont{Mousseau}} \bibnamefont{and}
  \bibinfo{author}{\bibfnamefont{G.~T.} \bibnamefont{Barkema}},
  \bibinfo{journal}{J. Phys.: Condens. Matter} \textbf{\bibinfo{volume}{16}},
  \bibinfo{pages}{S5183} (\bibinfo{year}{2004}).

\bibitem[{\citenamefont{Stillinger}(1988)}]{Sti88a}
\bibinfo{author}{\bibfnamefont{F.~H.} \bibnamefont{Stillinger}},
  \bibinfo{journal}{J. Chem. Phys.} \textbf{\bibinfo{volume}{88}},
  \bibinfo{pages}{7818} (\bibinfo{year}{1988}).

\bibitem[{\citenamefont{Stillinger}(1995)}]{Sti95a}
\bibinfo{author}{\bibfnamefont{F.~H.} \bibnamefont{Stillinger}},
  \bibinfo{journal}{Science} \textbf{\bibinfo{volume}{267}},
  \bibinfo{pages}{1935} (\bibinfo{year}{1995}).

\bibitem[{\citenamefont{Sciortino}(2005)}]{Stathmech}
\bibinfo{author}{\bibfnamefont{F.}~\bibnamefont{Sciortino}},
  \bibinfo{journal}{J. Stat. Mech.} p. \bibinfo{pages}{P05015}
  (\bibinfo{year}{2005}).

\bibitem[{\citenamefont{Horbach and Kob}(1999)}]{Horbach:1999ib}
\bibinfo{author}{\bibfnamefont{J.}~\bibnamefont{Horbach}} \bibnamefont{and}
  \bibinfo{author}{\bibfnamefont{W.}~\bibnamefont{Kob}},
  \bibinfo{journal}{Phys. Rev. B} \textbf{\bibinfo{volume}{60}},
  \bibinfo{pages}{3169} (\bibinfo{year}{1999}).

\bibitem[{\citenamefont{Saika-Voivod et~al.}(2004)\citenamefont{Saika-Voivod,
  Sciortino, and Poole}}]{SaikaVoivod:2004jf}
\bibinfo{author}{\bibfnamefont{I.}~\bibnamefont{Saika-Voivod}},
  \bibinfo{author}{\bibfnamefont{F.}~\bibnamefont{Sciortino}},
  \bibnamefont{and} \bibinfo{author}{\bibfnamefont{P.~H.} \bibnamefont{Poole}},
  \bibinfo{journal}{Phys. Rev. E} \textbf{\bibinfo{volume}{69}},
  \bibinfo{pages}{041503} (\bibinfo{year}{2004}).

\bibitem[{\citenamefont{Saika-Voivod et~al.}(2005)\citenamefont{Saika-Voivod,
  Sciortino, Grande, and Poole}}]{SaikaVoivod:2005vt}
\bibinfo{author}{\bibfnamefont{I.}~\bibnamefont{Saika-Voivod}},
  \bibinfo{author}{\bibfnamefont{F.}~\bibnamefont{Sciortino}},
  \bibinfo{author}{\bibfnamefont{T.}~\bibnamefont{Grande}}, \bibnamefont{and}
  \bibinfo{author}{\bibfnamefont{P.~H.} \bibnamefont{Poole}},
  \bibinfo{journal}{Phil. T. Roy. Soc. A} \textbf{\bibinfo{volume}{363}},
  \bibinfo{pages}{525} (\bibinfo{year}{2005}).

\bibitem[{\citenamefont{Berendsen et~al.}(1995)\citenamefont{Berendsen, van~der
  Spoel, and van Druren}}]{Berendsen:1995tn}
\bibinfo{author}{\bibfnamefont{H.~J.~C.} \bibnamefont{Berendsen}},
  \bibinfo{author}{\bibfnamefont{D.}~\bibnamefont{van~der Spoel}},
  \bibnamefont{and} \bibinfo{author}{\bibfnamefont{R.}~\bibnamefont{van
  Druren}}, \bibinfo{journal}{Comput. Phys. Commun.}
  \textbf{\bibinfo{volume}{91}}, \bibinfo{pages}{43} (\bibinfo{year}{1995}).

\bibitem[{\citenamefont{Lindahl et~al.}(2001)\citenamefont{Lindahl, Hess, and
  van~der Spoel}}]{Lindahl:2001bm}
\bibinfo{author}{\bibfnamefont{E.}~\bibnamefont{Lindahl}},
  \bibinfo{author}{\bibfnamefont{B.}~\bibnamefont{Hess}}, \bibnamefont{and}
  \bibinfo{author}{\bibfnamefont{D.}~\bibnamefont{van~der Spoel}},
  \bibinfo{journal}{J. Mol. Model.} \textbf{\bibinfo{volume}{7}},
  \bibinfo{pages}{306} (\bibinfo{year}{2001}).

\bibitem[{\citenamefont{van~der Spoel et~al.}(2005)\citenamefont{van~der Spoel,
  Lindahl, Hess, Groenhof, Mark, and Berendsen}}]{vanderSpoel:2005hz}
\bibinfo{author}{\bibfnamefont{D.}~\bibnamefont{van~der Spoel}},
  \bibinfo{author}{\bibfnamefont{E.}~\bibnamefont{Lindahl}},
  \bibinfo{author}{\bibfnamefont{B.}~\bibnamefont{Hess}},
  \bibinfo{author}{\bibfnamefont{G.}~\bibnamefont{Groenhof}},
  \bibinfo{author}{\bibfnamefont{A.~E.} \bibnamefont{Mark}}, \bibnamefont{and}
  \bibinfo{author}{\bibfnamefont{H.~J.~C.} \bibnamefont{Berendsen}},
  \bibinfo{journal}{J. Comput. Chem.} \textbf{\bibinfo{volume}{26}},
  \bibinfo{pages}{1701} (\bibinfo{year}{2005}).

\bibitem[{\citenamefont{Hess et~al.}(2008)\citenamefont{Hess, Kutzner, van~der
  Spoel, and Lindahl}}]{Hess:2008db}
\bibinfo{author}{\bibfnamefont{B.}~\bibnamefont{Hess}},
  \bibinfo{author}{\bibfnamefont{C.}~\bibnamefont{Kutzner}},
  \bibinfo{author}{\bibfnamefont{D.}~\bibnamefont{van~der Spoel}},
  \bibnamefont{and} \bibinfo{author}{\bibfnamefont{E.}~\bibnamefont{Lindahl}},
  \bibinfo{journal}{J. Chem. Theory Comput.} \textbf{\bibinfo{volume}{4}},
  \bibinfo{pages}{435} (\bibinfo{year}{2008}).

\bibitem[{\citenamefont{Abascal and Vega}(2010)}]{Abascal:2010dw}
\bibinfo{author}{\bibfnamefont{J.~L.~F.} \bibnamefont{Abascal}}
  \bibnamefont{and} \bibinfo{author}{\bibfnamefont{C.}~\bibnamefont{Vega}},
  \bibinfo{journal}{J. Chem. Phys.} \textbf{\bibinfo{volume}{133}},
  \bibinfo{pages}{234502} (\bibinfo{year}{2010}).

\bibitem[{\citenamefont{Sciortino et~al.}(2011)\citenamefont{Sciortino,
  Saika-Voivod, and Poole}}]{Sciortino:2011gf}
\bibinfo{author}{\bibfnamefont{F.}~\bibnamefont{Sciortino}},
  \bibinfo{author}{\bibfnamefont{I.}~\bibnamefont{Saika-Voivod}},
  \bibnamefont{and} \bibinfo{author}{\bibfnamefont{P.~H.} \bibnamefont{Poole}},
  \bibinfo{journal}{Phys. Chem.} \textbf{\bibinfo{volume}{13}},
  \bibinfo{pages}{19759} (\bibinfo{year}{2011}).

\bibitem[{\citenamefont{Vasisht et~al.}(2011)\citenamefont{Vasisht, Saw, and
  Sastry}}]{Vasisht:2011ch}
\bibinfo{author}{\bibfnamefont{V.~V.} \bibnamefont{Vasisht}},
  \bibinfo{author}{\bibfnamefont{S.}~\bibnamefont{Saw}}, \bibnamefont{and}
  \bibinfo{author}{\bibfnamefont{S.}~\bibnamefont{Sastry}},
  \bibinfo{journal}{Nat. Phys.} \textbf{\bibinfo{volume}{7}},
  \bibinfo{pages}{549} (\bibinfo{year}{2011}).

\bibitem[{\citenamefont{Yuan and Cormack}(2002)}]{Yuan:2002tt}
\bibinfo{author}{\bibfnamefont{X.}~\bibnamefont{Yuan}} \bibnamefont{and}
  \bibinfo{author}{\bibfnamefont{A.~N.} \bibnamefont{Cormack}},
  \bibinfo{journal}{Comp. Mater. Sci.} \textbf{\bibinfo{volume}{24}},
  \bibinfo{pages}{343} (\bibinfo{year}{2002}).

\bibitem[{\citenamefont{Elliott}(1991)}]{Elliott:2001tp}
\bibinfo{author}{\bibfnamefont{S.~R.} \bibnamefont{Elliott}},
  \bibinfo{journal}{Nature} \textbf{\bibinfo{volume}{354}},
  \bibinfo{pages}{445} (\bibinfo{year}{1991}).

\bibitem[{\citenamefont{Xie et~al.}(2013)\citenamefont{Xie, Long, Weigand,
  Moss, Carvalho, Roorda, Hejna, Torquato, and Steinhardt}}]{torquatopnas}
\bibinfo{author}{\bibfnamefont{R.}~\bibnamefont{Xie}},
  \bibinfo{author}{\bibfnamefont{G.~G.} \bibnamefont{Long}},
  \bibinfo{author}{\bibfnamefont{S.~J.} \bibnamefont{Weigand}},
  \bibinfo{author}{\bibfnamefont{S.~C.} \bibnamefont{Moss}},
  \bibinfo{author}{\bibfnamefont{T.}~\bibnamefont{Carvalho}},
  \bibinfo{author}{\bibfnamefont{S.}~\bibnamefont{Roorda}},
  \bibinfo{author}{\bibfnamefont{M.}~\bibnamefont{Hejna}},
  \bibinfo{author}{\bibfnamefont{S.}~\bibnamefont{Torquato}}, \bibnamefont{and}
  \bibinfo{author}{\bibfnamefont{P.~J.} \bibnamefont{Steinhardt}},
  \bibinfo{journal}{P. Natl. Acad. Sci. USA} \textbf{\bibinfo{volume}{110}},
  \bibinfo{pages}{13250} (\bibinfo{year}{2013}).

\bibitem[{\citenamefont{de~Graff and Thorpe}(2010)}]{thorpe}
\bibinfo{author}{\bibfnamefont{A.~M.~R.} \bibnamefont{de~Graff}}
  \bibnamefont{and} \bibinfo{author}{\bibfnamefont{M.~F.}
  \bibnamefont{Thorpe}}, \bibinfo{journal}{Acta Crystallogr A}
  \textbf{\bibinfo{volume}{66}}, \bibinfo{pages}{22} (\bibinfo{year}{2010}).

\bibitem[{\citenamefont{De~Michele
  et~al.}(2006{\natexlab{b}})\citenamefont{De~Michele, Tartaglia, and
  Sciortino}}]{DeMichele:2006iz}
\bibinfo{author}{\bibfnamefont{C.}~\bibnamefont{De~Michele}},
  \bibinfo{author}{\bibfnamefont{P.}~\bibnamefont{Tartaglia}},
  \bibnamefont{and}
  \bibinfo{author}{\bibfnamefont{F.}~\bibnamefont{Sciortino}},
  \bibinfo{journal}{J. Chem. Phys.} \textbf{\bibinfo{volume}{125}},
  \bibinfo{pages}{204710} (\bibinfo{year}{2006}{\natexlab{b}}).

\bibitem[{\citenamefont{Yuan and Cormack}(2003)}]{Yuan:2003es}
\bibinfo{author}{\bibfnamefont{X.}~\bibnamefont{Yuan}} \bibnamefont{and}
  \bibinfo{author}{\bibfnamefont{A.~N.} \bibnamefont{Cormack}},
  \bibinfo{journal}{J. of Non-Cryst. Solids} \textbf{\bibinfo{volume}{319}},
  \bibinfo{pages}{31} (\bibinfo{year}{2003}).

\bibitem[{\citenamefont{Poole et~al.}(2011)\citenamefont{Poole, Becker,
  Sciortino, and Starr}}]{Poole:2011bb}
\bibinfo{author}{\bibfnamefont{P.~H.} \bibnamefont{Poole}},
  \bibinfo{author}{\bibfnamefont{S.~R.} \bibnamefont{Becker}},
  \bibinfo{author}{\bibfnamefont{F.}~\bibnamefont{Sciortino}},
  \bibnamefont{and} \bibinfo{author}{\bibfnamefont{F.~W.} \bibnamefont{Starr}},
  \bibinfo{journal}{J. Phys. Chem. B} \textbf{\bibinfo{volume}{115}},
  \bibinfo{pages}{14176} (\bibinfo{year}{2011}).

\bibitem[{\citenamefont{Rahman and Stillinger}(1972)}]{dip1}
\bibinfo{author}{\bibfnamefont{A.}~\bibnamefont{Rahman}} \bibnamefont{and}
  \bibinfo{author}{\bibfnamefont{F.~H.} \bibnamefont{Stillinger}},
  \bibinfo{journal}{The Journal of Chemical Physics}
  \textbf{\bibinfo{volume}{57}}, \bibinfo{pages}{4009} (\bibinfo{year}{1972}).

\bibitem[{\citenamefont{Buch et~al.}(1998)\citenamefont{Buch, Sandler, and
  Sadlej}}]{dip2}
\bibinfo{author}{\bibfnamefont{V.}~\bibnamefont{Buch}},
  \bibinfo{author}{\bibfnamefont{P.}~\bibnamefont{Sandler}}, \bibnamefont{and}
  \bibinfo{author}{\bibfnamefont{J.}~\bibnamefont{Sadlej}},
  \bibinfo{journal}{The Journal of Physical Chemistry B}
  \textbf{\bibinfo{volume}{102}}, \bibinfo{pages}{8641} (\bibinfo{year}{1998}).

\bibitem[{\citenamefont{Ford et~al.}(2004)\citenamefont{Ford, Auerbach, and
  Monson}}]{monson}
\bibinfo{author}{\bibfnamefont{M.~H.} \bibnamefont{Ford}},
  \bibinfo{author}{\bibfnamefont{S.~M.} \bibnamefont{Auerbach}},
  \bibnamefont{and} \bibinfo{author}{\bibfnamefont{P.~A.}
  \bibnamefont{Monson}}, \bibinfo{journal}{J. Chem. Phys.}
  \textbf{\bibinfo{volume}{121}}, \bibinfo{pages}{8415} (\bibinfo{year}{2004}).

\bibitem[{\citenamefont{Bianchi et~al.}(2008)\citenamefont{Bianchi, Tartaglia,
  and Sciortino}}]{bianchisilica}
\bibinfo{author}{\bibfnamefont{E.}~\bibnamefont{Bianchi}},
  \bibinfo{author}{\bibfnamefont{P.}~\bibnamefont{Tartaglia}},
  \bibnamefont{and}
  \bibinfo{author}{\bibfnamefont{F.}~\bibnamefont{Sciortino}},
  \bibinfo{journal}{J. Chem. Phys.} \textbf{\bibinfo{volume}{129}},
  \bibinfo{pages}{224904} (\bibinfo{year}{2008}).

\bibitem[{\citenamefont{Saika-Voivod
  et~al.}(2011{\natexlab{b}})\citenamefont{Saika-Voivod, King, Tartaglia,
  Sciortino, and Zaccarelli}}]{SaikaVoivod:2011by}
\bibinfo{author}{\bibfnamefont{I.}~\bibnamefont{Saika-Voivod}},
  \bibinfo{author}{\bibfnamefont{H.~M.} \bibnamefont{King}},
  \bibinfo{author}{\bibfnamefont{P.}~\bibnamefont{Tartaglia}},
  \bibinfo{author}{\bibfnamefont{F.}~\bibnamefont{Sciortino}},
  \bibnamefont{and}
  \bibinfo{author}{\bibfnamefont{E.}~\bibnamefont{Zaccarelli}},
  \bibinfo{journal}{J. Phys.: Condens. Matter} \textbf{\bibinfo{volume}{23}},
  \bibinfo{pages}{285101} (\bibinfo{year}{2011}{\natexlab{b}}).

\bibitem[{\citenamefont{Sciortino}(2008)}]{statphys}
\bibinfo{author}{\bibfnamefont{F.}~\bibnamefont{Sciortino}},
  \bibinfo{journal}{Eur. Phys. J. B} \textbf{\bibinfo{volume}{64}},
  \bibinfo{pages}{505} (\bibinfo{year}{2008}).

\bibitem[{\citenamefont{Li et~al.}(2004)\citenamefont{Li, Tseng, Kwon,
  d'Espaux, Bunch, McEuen, and Luo}}]{luoxy}
\bibinfo{author}{\bibfnamefont{Y.}~\bibnamefont{Li}},
  \bibinfo{author}{\bibfnamefont{Y.~D.} \bibnamefont{Tseng}},
  \bibinfo{author}{\bibfnamefont{S.~Y.} \bibnamefont{Kwon}},
  \bibinfo{author}{\bibfnamefont{L.}~\bibnamefont{d'Espaux}},
  \bibinfo{author}{\bibfnamefont{J.~S.} \bibnamefont{Bunch}},
  \bibinfo{author}{\bibfnamefont{P.~L.} \bibnamefont{McEuen}},
  \bibnamefont{and} \bibinfo{author}{\bibfnamefont{D.}~\bibnamefont{Luo}},
  \bibinfo{journal}{Nat. Mater.} \textbf{\bibinfo{volume}{3}},
  \bibinfo{pages}{38} (\bibinfo{year}{2004}).

\bibitem[{\citenamefont{Montarnal et~al.}(2011)\citenamefont{Montarnal,
  Capelot, Tournilhac, and Leibler}}]{vitrimers}
\bibinfo{author}{\bibfnamefont{D.}~\bibnamefont{Montarnal}},
  \bibinfo{author}{\bibfnamefont{M.}~\bibnamefont{Capelot}},
  \bibinfo{author}{\bibfnamefont{F.}~\bibnamefont{Tournilhac}},
  \bibnamefont{and} \bibinfo{author}{\bibfnamefont{L.}~\bibnamefont{Leibler}},
  \bibinfo{journal}{Science} \textbf{\bibinfo{volume}{334}},
  \bibinfo{pages}{965} (\bibinfo{year}{2011}).

\bibitem[{\citenamefont{Zhang et~al.}(2005)\citenamefont{Zhang, Wang, and
  M{\"o}hwald}}]{mohovald}
\bibinfo{author}{\bibfnamefont{G.}~\bibnamefont{Zhang}},
  \bibinfo{author}{\bibfnamefont{D.}~\bibnamefont{Wang}}, \bibnamefont{and}
  \bibinfo{author}{\bibfnamefont{H.}~\bibnamefont{M{\"o}hwald}},
  \bibinfo{journal}{Angew. Chem. Int. Ed.} \textbf{\bibinfo{volume}{44}},
  \bibinfo{pages}{7767} (\bibinfo{year}{2005}).

\end{thebibliography}

\end{document}